\documentclass[10pt,twocolumn,aps,pra,showpacs,nofootinbib,superscriptaddress]{revtex4-1}\usepackage[english]{babel} 
\usepackage[usenames, dvipsnames]{color} 
\usepackage{graphicx} 
\usepackage{bm} 
\usepackage{amsmath}
\usepackage{amsthm} 
\usepackage{amssymb} 
\usepackage{times} 
\usepackage{mathrsfs}
\usepackage[pdftex,colorlinks=true,urlcolor=blue,linkcolor=Blue,citecolor=RedViolet]{hyperref} 
\usepackage{appendix}
\newcommand{\hy}[1]{\hyperlink{#1}{\color{Gray} #1}}%my hyperlink
\begin{document}
\title{Weyl-Wigner Representation of Canonical Equilibrium States}
\author{F. Nicacio}
\email{nicacio@if.ufrj.br} 
\affiliation{Instituto de F\'isica, Universidade Federal do Rio de Janeiro, 
             21941-972, RJ, Brazil.}
\affiliation{ 
              Universit\"at Wien, NuHAG, Fakult\"at f\"ur Mathematik, 
              A-1090 Wien, Austria.                                   }             
%%%%%%%%%%%%%%%%%%%%%%%%%%%%%%%%%%%%%%%%%%%%%%%%%%%%%%%%%%%%%%%%%%%%%%%%%%%%%%%%%%%%%%%%%
\begin{abstract}
\noindent The Weyl-Wigner representations for canonical thermal equilibrium 
quantum states are obtained for the whole class of quadratic Hamiltonians through a 
Wick rotation of the Weyl-Wigner symbols of Heisenberg and metaplectic operators. 
The behavior of classical structures inherently associated to these unitaries  
is described under the Wick mapping,  
unveiling that a thermal equilibrium state is fully determined by 
a complex symplectic matrix,  
which sets all of its thermodynamical properties. 
The four categories of Hamiltonian dynamics
(Parabolic, Elliptic, Hyperbolic and Loxodromic) are analyzed. 
Semiclassical and high temperature approximations are derived and compared 
to the classical and/or quadratic behavior.  
\end{abstract}
%%%%%%%%%%%%%%%%%%%%%%%%%%%%%%%%%%%%%%%%%%%%%%%%%%%%%%%%%%%%%%%%%%%%%%%%%%%%%%%%%%%%%%%%%
\maketitle
%%%%%%%%%%%%%%%%%%%%%%%%%%%%%%%%%%%%%%%%%%%%%%%%%%%%%%%%%%%%%%%%%%%%%%%%%%%%%%%%%%%%%%%%%

%%%%%%%%%%%%%%%%%%%%%%%%%%%%%%%%%%%%%%%%%%%%%%%%%%%%%%%%%%%%%%%%%%%%%%%%%%%%%%%%%%%%%%%%%
%%%%%%%%%%%%%%%%%%%%%%%%%       INTRODUCTION     %%%%%%%%%%%%%%%%%%%%%%%%%%%%%%%%%%%%%%%%
%%%%%%%%%%%%%%%%%%%%%%%%%%%%%%%%%%%%%%%%%%%%%%%%%%%%%%%%%%%%%%%%%%%%%%%%%%%%%%%%%%%%%%%%%
Following the statistical physics postulates \cite{statphys},  
the state of a system in equilibrium with a canonical thermal reservoir is described by
the density operator 
\begin{equation}                                                                         \label{Eq:tstate}
\hat \rho_\text{T} = \frac{{\rm e}^{-\beta \hat H}}{\mathcal Z_{\beta}}, \,\,\, 
\mathcal Z_{\beta} := {\rm Tr}\, {\rm e}^{-\beta \hat H}, 
\end{equation}
where $\beta:= (k_{\rm B} T)^{-1} \in \mathbb R$ is the ``inverse temperature'', 
$k_{\rm B}$ is the Boltzmann constant and $\hat H$ is the Hamiltonian of the system. 
The partition function (\hypertarget{PF}{PF}) $\mathcal Z_{\beta}$ provides the normalization of the state 
and is a central object of the theory, 
since it is the fist step towards the derivation of thermodynamical 
function and potentials \cite{statphys}.     

The evolution of a (time independent) quantum system is performed by 
the unitary operator 
\begin{equation}                                                                         \label{Eq:UnitOp}
\hat U_t := {\rm e}^{-i \hat H t/\hbar },
\end{equation}
which can be related to the thermal state in (\ref{Eq:tstate}) through
\begin{equation}                                                                         \label{Eq:WickRot}
t \mapsto -i\hbar\beta.  
\end{equation}
This simple holomorphic mapping, known as {\it Wick rotation} \cite{wick}, 
is particularly useful and constitutes basic tool 
for quantum field theory, 
see for instance \cite{peskin,zee}, 
where propagators (path-integrations in Minkowski space) 
are mapped into Euclidean path integrals. 
Surprisingly enough, 
hardly one will find this subject in statistical mechanics textbooks \cite{Note1}, 
and only two examples, the free particle and the harmonic oscillator, 
are presented in the standard quantum mechanics literature \cite{QuantMech}.

The mentioned examples 
are embraced by the wider category of the systems described by 
quadratic Hamiltonians (\hypertarget{QH}{QH}), 
which constitute the basic building 
blocks for the study of conservative dynamical systems 
in classical and in quantum mechanics \cite{ozoriobook,gutzwiller}. 
For this category, 
a group theoretical approach elegantly combines classical 
and quantum mechanics over a phase-space background supplied by 
the Weyl-Wigner-Moyal description of quantum mechanics 
\cite{yeh,ozorio1998,littlejohn1986,gossonbook2006}. 

For generic quantum dynamics, 
semiclassical approximations are useful methods to 
describe the system behavior when the constant $\hbar$ 
is very small when compared to a characteristic action,
which roughly constitutes the limit $\hbar \rightarrow 0$ \cite{ozoriobook,gutzwiller}. 
In this limit, inherent classical structures emerge, 
{\it e.g.}, 
the famous WKB method shows that 
the phase of the wave function is governed by 
a classical Hamilton-Jacobi equation \cite{ozoriobook,gutzwiller}.   
Quadratic Hamiltonians provides, by one side, 
the best known examples of application for semiclassical methods  
in what concerns the quantum-classical correspondence \cite{ozoriobook,gutzwiller}. 
By another side, some of the semiclassical techniques 
are exact for such kind of Hamiltonians, for instance,  
the Moyal bracket collapses into 
the classical Poisson bracket \cite{moyal}, while  
for any other non-trivial Hamiltonian, 
it constitutes an expansion in powers of $\hbar$.  

Quadratic Hamiltonians are 
the commonly realizable operations in optics laboratories  
for the manipulation of the continuous degrees of freedom (quadratures)
of electromagnetic field \cite{QuantOptics}.
Nowadays, these powerful techniques are also devoted 
to encoding, manipulate, transport, and store information
by quantum protocols associated to 
continuous degrees of freedom states \cite{CVQuantInfo}. 
Thermal states of the electromagnetic field occupies a privileged 
position at this scenario, 
due to a lack of a Weyl-Wigner description (or any other equivalent) 
for all thermal states,  
distinct theoretical methods were developed just to determine
ensemble averages at non-zero temperatures \cite{takahashi} 
and the Wigner function itself for some 
\hy{QH} thermal states \cite{hong}. 

In the scope of open quantum system dynamics,  
the Markovian interaction of continuous-variable quantum system
with an external and uncontrollable environment
can lead the system to a steady-state \cite{MarkOpenSys}. 
Theoretically, thermal equilibrium states (\ref{Eq:tstate}) 
associated to \hy{QH}s can be generated by 
an appropriate environmental interaction \cite{nicacio11}.        
The robustness of the steady-state, 
since it does not depend on the initial state, 
but only on the environment,  
is a valuable tool for state engineering, stabilization and design, 
as detailed in \cite{nicacio11} and the references therein.     
 
In this work, the Wigner-Weyl symbols 
(and thus, the Wigner and the characteristic function) of canonical 
thermal equilibrium states will be determined for the whole class of systems 
described by \hy{QH}s with a generic number of degrees of freedom.  
The derivation is performed applying a Wick rotation to the Weyl-Wigner symbols 
associated to groups of unitary operators, named Heisenberg and Metaplectic. 
These operators are quantum representations of 
classical translations and (real) symplectic transformations. 
The generality of the results obtained for any \hy{QH} 
is due to the duality between the Wigner and the Weyl representations, 
where an unavoidable divergence of one is compensated by the well behavior of the other.  

The set of thermal states is shown to be completely described by  
a complex symplectic group raised by a Wick rotation of 
the classical phase-space. Interesting enough, 
the categorization of the four types of classical symplectic dynamics 
(Parabolic, Elliptic, Hyperbolic and Loxodromic) 
is extended to the thermal states and examples are given.  
The Elliptic case corresponds to the class of 
positive-definite \hy{QH}, 
which includes the harmonic oscillator system; 
it is the only case where both Wigner-Weyl symbols are Gaussians 
and has been extensively studied in the literature, 
see for instance \cite{gaussian}. 
The inherent covariance of symbols under linear canonical transformations
when applied to thermal states does not change thermodynamical 
properties of each category.  
Limits on temperature and on $\hbar$ show connections between 
the quantum and classical thermal states. 

This work is organized as follows. 
Section~\ref{Sec:WWF} begins with the Wigner-Weyl formalism description 
and finishes with the Wick rotation of the unitary symbols.
The structure of the classical canonical transformations 
and their Wick rotated version are placed in Sec.\ref{Sec:se}.
The unitary subgroups related to the classical canonical transformations 
are described in Sec.\ref{Sec:sue}, while 
the symbols for the thermal equilibrium states and their 
properties are calculated and determined in Sec.\ref{Sec:toqh}.      
Approximations for non-\hy{QH}s 
are in Sec.\ref{Sec:gh}. 
In Sec.\ref{Sec:Ex}, 
several examples of thermal states generated by 
\hy{QH}s are given and its properties analyzed according to 
the four categories of symplectic matrices.
Finally, the conclusions and perspectives are presented in Sec.\ref{Sec:conc}.  

\section{Weyl-Wigner Formalism}                                   \label{Sec:WWF}  %%%%%%
%%%%%%%%%%%%%%%%%%%%%%%%%%%%%%%%%%%%%%%%%%%%%%%%%%%%%%%%%%%%%%%%%%%%%%%%%%%%%%%%%%%%%%%%%
%%%%%%%%%%%%%%%%%%%%%%%%%%%%%%%%%%%%%%%%%%%%%%%%%%%%%%%%%%%%%%%%%%%%%%%%%%%%%%%%%%%%%%%%%
Consider a quantum systems described by $n$ continuous bosonic degrees of freedom. 
The generalized coordinates 
$\hat q := (\hat q_1,...,\hat q_n)^\dagger$ 
together with the canonical conjugated momenta 
$\hat p := (\hat p_1,...\hat p_n)^\dag$  
are written collectively as $2n$-column-vector:
$%begin{equation}                                                                         \label{vecx}
\hat x := (\hat q_1,...,\hat q_n, \hat p_1,...\hat p_n)^\dag
$. %end{equation}
In this notation, the canonical commutation relation (\hypertarget{CCR}{CCR})
is written compactly as $[\hat x_j , \hat x_k ] = i \hbar \, \mathsf J_{jk}$ 
with ${\sf J}_{jk}$ given by the elements of the symplectic matrix 
\begin{equation}                                                                         \label{Eq:SympJ}
 \mathsf J := 
\left(  
\begin{array}{cc} 
       {\bf 0}_n   & \mathsf I_n  \\
      -\mathsf I_n & {\bf 0}_n         
       \end{array}
 \right) = - \mathsf J^\top = - \mathsf J^{-1}.
\end{equation}

The Weyl translation (or the Heisenberg operator) 
is an unitary operator defined as \cite{ozorio1998}
\begin{equation}                                                                         \label{Eq:HeisOp}
\hat T_\xi := {\exp}\!\left[- \frac{i}{\hbar} \hat x \wedge \xi  \right], 
\,\,\, \hat x \wedge \xi := \mathsf J \hat x \cdot \xi,    
\end{equation}
where the column vector  
$\xi := (\xi_{q_1},...,\xi_{q_n},\xi_{p_1},...,\xi_{p_n}){\!^\top} \in \mathbb{R}^{2n}$,  
sets the direction of a translation of the operator $\hat x$, 
{\it i.e.}, 
$\hat T^\dag_{\xi}\hat x \hat T_{\xi}=\hat x+\xi\hat 1$ 
with 
$\hat T_\xi^{-1}=\hat T_\xi^\dagger=\hat T_{-\xi}$.

The parity operator \cite{QuantMech} will be denoted 
$\hat R_0$ and its action is 
described by $\hat R_0^\dagger \hat x \hat R_0 = - \hat x$. 
It is an involutory operator, since $\hat R_0^2 = \hat 1$, 
and thus $\hat R_0 = \hat R_0^\dagger = \hat R_0^{-1}$. 
The reflection operator is defined as \cite{ozorio1998} 
\begin{eqnarray}                                                                         \label{Eq:ReflOp}
\hat R_x := \hat T_x \hat R_0 \hat T_x^\dagger = \hat R_x^\dagger = \hat R_x^{-1},  
\end{eqnarray}
where $x := (q_1,...,q_n,p_1,...,p_n)^\top \in \mathbb R^{2n}$ 
is a column vector indicating the reflection point, {\it i.e.}, 
$\hat R_x \hat x \hat R_x=-\hat x+2x\hat 1$.  
% It is possible to show that $\hat R_x$ is a 
% symplectic Fourier transform of $\hat T_\xi$, namely
% \begin{eqnarray}                                                                         \label{Eq:ReflOp2}
% \hat R_x := \int \!\!\frac{{\rm d}^{2n}\xi}{(4\pi\hbar)^n} \,
% {\exp}\!\left[\frac{i}{\hbar}  \hat x \wedge \xi \right]\, \hat T_\xi,  
% \end{eqnarray}

%
Both sets of translations and of reflections constitute a basis for the 
vector space of the operators acting on the Hilbert space of a 
continuous variable quantum system, {\it i.e.}, 
an arbitrary operator $\hat A$ can be uniquely expanded as 
\begin{equation}                                                                         \label{Eq:WWreps}
\hat A = \int \!\!\frac{{\rm d}^{2n}\xi }{(2\pi\hbar)^n} \, \tilde{A}(\xi) \,  \hat T_\xi   
       = \int \!\!\frac{{\rm d}^{2n}  x }{(\pi\hbar )^n} \, A(x) \, {\hat R}_x,               
\end{equation}
which are, respectively, 
the Weyl and the Wigner representations of $\hat A$ \cite{ozorio1998}. 
The coefficients $\tilde{A}(\xi)$ and $A(x)$ are, respectively, 
defined through the inner products 
\begin{equation}                                                                         \label{Eq:WWsymbs}
\tilde{A}(\xi) = \textrm{Tr}\,(\hat A \hat T_{\xi}^\dag), \,\,\,\,     
A(x)   = 2^n \textrm{Tr}\,( \hat A {\hat R}_x),
\end{equation}
by virtue of \cite{ozorio1998}
\begin{equation}                                                                         \label{Eq:TRtrace}
\text{Tr}\,(\hat T_\xi \hat T^ \dag_{\xi'}) = 
2^{2n}\text{Tr}\,({\hat R}_{\xi}{\hat R}_{\xi'}) = (2\pi\hbar)^n\delta^{2n}(\xi-\xi').                               
\end{equation}
It is also important to mention that
\begin{equation}                                                                         \label{Eq:TRtrace2}
\text{Tr}\,(\hat T_\xi) =  (2\pi\hbar)^n\delta^{2n}(\xi), \,\,\, 
\text{Tr}\,({\hat R}_{\xi}) = 1/2^{n}.                                
\end{equation}

The coefficients in (\ref{Eq:WWsymbs}) are known, respectively, 
as the Weyl and Wigner symbols of the operator $\hat A$. 
The change of basis (\ref{Eq:ReflOp}) relates these symbols through 
a symplectic Fourier transform, {\it viz.}, 
\begin{equation}                                                                         \label{Eq:SympFour}                                                             
\tilde{A}(\xi) =  \int\!\! \frac{{\rm d}^{2n}x }{(2\pi\hbar)^n}\, A(x)\;
                 {\rm e}^{\frac{i}{\hbar}  x \wedge \xi }.                       
%
%A(x)&=&  \int \!\!  \frac{d\xi}{(2\pi\hbar)^n} \, \tilde{A}(\xi)\;
%                        {\rm e}^{\frac{i}{\hbar}x^{\!\top} {\sf J} \xi }.                \label{Weylexp}  
\end{equation}

In particular, the Wigner function $W(x)$ of a quantum state is (a normalized version of) 
the Wigner symbol associated with the corresponding density operator 
$\hat \rho$ \cite{parity,ozorio1998}, that is, 
\begin{equation}                                                                         \label{Eq:Wigfunc}
W(x) := \frac{1}{(\pi\hbar)^n} {\rm Tr} \left[ \hat \rho {\hat R}_x \right].  
\end{equation}
Its symplectic Fourier transform is the characteristic function (the Weyl symbol)
of $\hat \rho$ \cite{ozorio1998}:
\begin{equation}                                                                         \label{Eq:Charfunc}
\chi(\xi) \! = \!\! \int \!\!\! \frac{{\rm d}^{2n} x}{(2\pi\hbar)^n} \, W(x) \,
                              {\rm e}^{ \frac{i}{\hbar} x \wedge \xi }
             = \frac{1}{(2\pi\hbar)^n}  {\rm Tr} 
             \left[\hat \rho \hat T^{\dagger}_{\xi} \right].  
\end{equation}
%
%%%%%%%%%%%%%%%%%%%%%%%%%%%%%%%%%%%%%%%%%%%%%%%%%%%%%%%%%%%%%%%%%%%%%%%%%%%%%%%%%%%%%%%%%
\subsection*{Equilibrium States, Unitary Operators and Wick Rotation}
%%%%%%%%%%%%%%%%%%%%%%%%%%%%%%%%%%%%%%%%%%%%%%%%%%%%%%%%%%%%%%%%%%%%%%%%%%%%%%%%%%%%%%%%%
The Weyl-Wigner symbols for the canonical 
equilibrium state in (\ref{Eq:tstate}) are determined, respectively, 
by (\ref{Eq:Wigfunc}) and (\ref{Eq:Charfunc}). 
However, it is interesting to describe the symbols for the operator 
$\exp(-\beta \hat H)$ and the \hy{PF} as a functional of these symbols.  
Thus, considering the expansions in (\ref{Eq:WWreps}), 
the symbols (\ref{Eq:WWsymbs}) for the thermal operator are  
% \begin{equation}                                                                         \label{eq:WWexpEq}
% {\rm e}^{-\beta \hat H} = 
% \int \!\!\frac{{\rm d}^{2n}\xi }{(2\pi\hbar)^n} \, \tilde{E}_\beta(\xi) \,  \hat T_\xi   
%        = \int \!\!\frac{{\rm d}^{2n}  x }{(\pi\hbar )^n} \, E_\beta(x) \, {\hat R}_x,   
% \end{equation}
% where the Weyl and Wigner symbols are given by (\ref{Eq:WWsymbs}) 
\begin{equation}                                                                         \label{Eq:ThOpSymb}
\tilde{E}_\beta(\xi) := \textrm{Tr}( {\rm e}^{-\beta \hat H} \hat T_{\xi}^\dag), \,\,\,\,     
E_\beta(x) := 2^n \textrm{Tr}\,( {\rm e}^{-\beta \hat H} {\hat R}_x).
\end{equation}
Complex conjugating and taking into account that 
the thermal operator is Hermitian, 
above symbols (actually the symbols of any Hermitian operator) 
are such that 
\begin{equation}                                                                         \label{Eq:HermSymmet}
\tilde{E}_\beta(\xi) = [\tilde{E}_\beta(-\xi)]^\ast, \,\,\,\,     
E_\beta(x) =  [E_\beta(x)]^\ast.
\end{equation}

Taking the trace of (\ref{Eq:WWreps}) and using (\ref{Eq:TRtrace2}), 
the expression for the \hy{PF} of the thermal state 
as functionals of the symbols becomes
\begin{equation}                                                                         \label{Eq:Partfunc}
\mathcal Z_{\beta} := {\rm Tr}\,{ \rm e}^{-\beta \hat H} = 
\tilde{E}_\beta(0) = 
\frac{1}{(2\pi\hbar )^n} \int\!\! {\rm d}^{2n}x \, E_\beta(x),      
\end{equation}
where the second equality is a manifestation of the in\-de\-pen\-dence of the trace 
on a specific basis.   

Joining (\ref{Eq:ThOpSymb}) and (\ref{Eq:Partfunc}), 
the characteristic and Wigner functions of the thermal state (\ref{Eq:tstate}) 
are 
\begin{equation}                                                                         \label{Eq:ThWigChFunc}
\chi(\xi) = \frac{\tilde{E}_\beta(\xi)}{(2\pi\hbar)^n\tilde{E}_\beta(0)}, 
\,\,\,
W(x) = \frac{E_\beta(x)}{\int\! {\rm d}^{2n}x \, E_\beta(x)} ,   
\end{equation}
as they should be from the definitions (\ref{Eq:Wigfunc}) and (\ref{Eq:Charfunc}).  

The expansions in (\ref{Eq:WWreps}) for the unitary operator in (\ref{Eq:UnitOp}) are  
% \[%\begin{equation}                                                                         \label{Eq:UnitExp}
% \hat U_t = 
% \int\!\! \frac{{\rm d}^{2n}\xi }{(2\pi\hbar)^n} \, \tilde{U}_t(\xi) \,  \hat T_\xi   
%        = \int\!\! \frac{{\rm d}^{2n}  x }{(\pi\hbar )^n} \, U_t(x)  \, {\hat R}_x,   
% \]%end{equation}
% %
% where the Weyl and Wigner symbols are given by (\ref{Eq:WWsymbs}) 
\begin{equation}                                                                         \label{Eq:UnitOpSymb}
\tilde{U}_t(\xi) = \textrm{Tr}( \hat U_t \hat T_{\xi}^\dag), \,\,\,\,     
{U}_t(x) = 2^n \textrm{Tr}( \hat U_t {\hat R}_x),
\end{equation}
and taking into account that $\hat U_t^\dagger = \hat U_{-t}$, 
above symbols are, respectively, such that
\begin{equation}                                                                         \label{Eq:UnitSymmet}
\tilde{U}_{-t}(\xi) = [\tilde{U}_{t}(-\xi)]^\ast, \,\,\,\,     
{U}_{-t}(x) = [{U}_{t}(x)]^\ast.
\end{equation}

% The Weyl and Wigner expansions (\ref{Eq:WWreps}) 
% show that translations or reflections 
% span the vector space of the operators over a Hilbert space, 
% which is the same as say that a

Since any operator can be expressed as a {\it linear} combination 
of translations or reflections, 
the Wick rotation (\ref{Eq:WickRot}) is 
readily applicable to the level of the coefficients of the expansion, {\it i.e.}, 
the Weyl-Wigner symbols. 
However, to take into account 
the nature of the operators at the symbols level, 
which are explicitly manifested in 
(\ref{Eq:HermSymmet}) and (\ref{Eq:UnitSymmet}), 
it is preferable to express the Wick mapping as  
\begin{equation}                                                                         \label{Eq:WickThSymb}
\begin{aligned}
\tilde{E}_\beta(\xi) &= \tfrac{1}{2} \tilde{U}_{-i \hbar \beta }(\xi) + 
                        \tfrac{1}{2} [\tilde{U}_{-i \hbar \beta }(-\xi)]^\ast,                      \\
E_\beta(x) &= % 
%              \tfrac{1}{2}  {U}_{-i \hbar \beta }(x) + 
%              \tfrac{1}{2} [{U}_{-i \hbar \beta}(x)]^\ast = 
              {\rm Re}\left[ {U}_{-i \hbar \beta }(x) \right].
\end{aligned}
\end{equation}
For the purposes of this work, it is sufficient to consider 
$\tilde{E}_\beta(\xi) = \tilde{U}_{-i \hbar \beta }(\xi)$ and
$E_\beta(x) = {U}_{-i \hbar \beta }(x)$
instead of above equations, however Eq.(\ref{Eq:WickThSymb}) 
are suitable for generalizations, 
for instance, in the case of time-dependent Hamiltonians.   
The inverse mapping is
\[
\tilde{U}_{t}(\xi) = \tilde{E}_{{it}/{\hbar}}(\xi),   \,\,\,
{U}_{t}(x) = {E}_{{it}/{\hbar} }(x).
\]%\end{equation}

Even though the problem seems to be solved by Eq.(\ref{Eq:WickThSymb}), 
the obtainment of ${U}_{t}(x)$ or $\tilde{U}_{t}(\xi)$ analytically 
is rare and restricted to few cases. 
In general, one considers (semiclassical) approximations valid for limited 
time intervals \cite{ozorio1998,littlejohn1986}. 
Even for \hy{QH}s,  
the obtainment of the propagator 
(unitary operator expanded in a certain basis) 
is not a trivial question. 
Remember that a closed expression for 
the propagator of the Harmonic oscillator in quantum mechanics 
textbooks relies on the celebrated, as well as intricate, 
Mehler formula for a sum of products of Hermite polynomials \cite{QuantMech}.  

Fortunately, the dynamics of quantum systems evolved by a 
generic \hy{QH} nowadays is elegantly developed 
though a theoretical group representation approach based 
on classical dynamics. 

%%%%%%%%%%%%%%%%%%%%%%%%%%%%%%%%%%%%%%%%%%%%%%%%%%%%%%%%%%%%%%%%%%%%%%%%%%%%%%%%%%%%%%%%%
%%%%%%%%%%%%%%%%%%%%%%%%%%%%%%%%%%%%%%%%%%%%%%%%%%%%%%%%%%%%%%%%%%%%%%%%%%%%%%%%%%%%%%%%%
\section{Symplectic Evolution}                                         \label{Sec:se}%%%%
%%%%%%%%%%%%%%%%%%%%%%%%%%%%%%%%%%%%%%%%%%%%%%%%%%%%%%%%%%%%%%%%%%%%%%%%%%%%%%%%%%%%%%%%%
%%%%%%%%%%%%%%%%%%%%%%%%%%%%%%%%%%%%%%%%%%%%%%%%%%%%%%%%%%%%%%%%%%%%%%%%%%%%%%%%%%%%%%%%%
% The symbol theory for unitary operators which will be translated for equilibrium states 
% is supported on classical Hamiltonian systems through a group theoretical approach, 
% {\it i.e.}, classical and quantum evolutions are faced as representations 
% of the same group for a given Hamiltonian. 
%
This section reviews some facts about classical \hy{QH}s
and also includes, at the end, a discussion about their Wick rotated version,
both of which will be necessary for the constructions of 
the Wigner-Weyl symbols. 

The simplest non-trivial evolution of a mechanical system is described 
by the linear Hamiltonian
\begin{equation}                                                                         \label{Eq:ClLinHam}
H_1 = x \wedge \zeta + H_0,  
\end{equation}
where $\zeta \in \mathbb R^{2n}$ is a constant column vector and  
$H_0 \in \mathbb R$ is a constant. 
Under $H_1(x)$, an initial condition in phase space $x_0 \in \mathbb R^{2n}$  
evolves to $\varphi_1(x_0,t) = x_0 + \zeta t$. 
The Hamiltonians (\ref{Eq:ClLinHam}) constitute the Heisenberg 
Lie algebra $\mathfrak{h}(2n)$ under the Poisson bracket operation.   

A ``new'' Hamiltonian 
$H'_1 = x \wedge \zeta' + H'_0$ generates the flux 
$\varphi'_1(x_0,t) = x_0 + \zeta' t$; 
since $\varphi'_1 \circ \varphi_1(x_0,t) = x_0 + (\zeta + \zeta') t$, 
the flux of these Hamiltonian are members of an additive Abelian group, 
sometimes also called the Heisenberg group of translations ${\rm H}(2n)$, 
generated by vectors $\mathsf J^\top \zeta \in \mathbb R^{2n}$.     

Consider now the classical \hy{QH} 
\begin{equation}                                                                         \label{Eq:ClQuadHam}
H_2 := \tfrac{1}{2} x \cdot \mathbf H x,    
\end{equation}
where $\mathbf H = \mathbf H^\top \in {\rm Mat}(2n,\mathbb R)$ is the 
Hessian matrix. 
The set of these Hamiltonians constitutes a Lie algebra 
(under the Poisson bracket) dubbed {\it metaplectic algebra} 
$\mathfrak{mp}(2n)$ \cite{littlejohn1986,gossonbook2006}. 
The phase-space flux of (\ref{Eq:ClQuadHam}) is $\varphi_2(x_0,t) = \mathsf S_t x_0$,  
where 
\begin{equation}                                                                         \label{Eq:SympEv}
\mathsf S_t := {\exp}[{\mathsf J {\bf H} t}]  
\end{equation}
is an element of the real symplectic group 
${\rm Sp}(2n,\mathbb R):= 
\{\mathsf M \in {\rm Mat}(2n, \mathbb  R) \, | \,  
\mathsf M^{\top} \mathsf J \mathsf M = \mathsf J \}$
for $\mathsf J$ in (\ref{Eq:SympJ}). 

Not all matrices ${\sf M} \in {\rm Sp}(2n,\mathbb R)$ can be written as (\ref{Eq:SympEv}), 
while matrices of the form (\ref{Eq:SympEv}) constitutes an uniparametric subgroup 
of ${\rm Sp}(2n,\mathbb R)$ for a given generator 
(also called Hamiltonian matrix)  
$\mathsf J {\bf H} \in \mathfrak{sp}(2n)$, where $\mathfrak{sp}(2n)$ 
is the symplectic Lie algebra. 
Any symplectic matrix ${\sf M} \in {\rm Sp}(2n,\mathbb R)$
is such that  $\det \mathsf M = 1$, 
since $\mathsf J = \mathsf M^\top \mathsf J \mathsf M$ implies \cite{gossonbook2006}
\begin{equation}                                                                         \label{Eq:sympPf}           
{\rm Pf}(\mathsf J) = {\rm Pf}(\mathsf M^\top \mathsf J \mathsf M) = 
(\det \mathsf M) \, {\rm Pf}(\mathsf J),  
 \end{equation}
where the second equality is a property of the Pfaffian \cite{jacobson}. 

The Hamiltonian nature of the generators 
$\mathsf J\mathbf H \in \mathfrak{sp}(2n)$
imposes strong constrains to the symplectic dynamics.  
The characteristic polynomial 
\[\begin{aligned}
P(\lambda) :
&= \det\left(\mathsf J\mathbf H -\lambda \mathsf I_{2n} \right) 
=  \det\left(\mathbf H \mathsf J^\top -\lambda \mathsf I_{2n} \right) \\
&= \det\left(\mathsf J^\top \mathbf H  -\lambda \mathsf I_{2n} \right) 
%=  \det\left(\mathsf J\mathbf H + \lambda \mathsf I_{2n} \right) 
= P\left(-\lambda\right),  
\end{aligned}\]
as shown, is an even function of $\lambda$.  
Furthermore, a complex eigenvalue always appears together 
with its complex conjugate, 
since $\det(\mathsf J\mathbf H) \in \mathbb R$.
These constrains show that the eigenvalues of a Hamiltonian matrix
$\mathsf J \bf H$ must fall into four cases \cite{arnold}: 
\begin{itemize}
\item[\hypertarget{(P)}{(P)}] 
Parabolic: a pair of null eigenvalues; \vspace{-0.25cm}
\item[\hypertarget{(H)}{(H)}] 
Hyperbolic: a pair $(k,-k)$ for $k \in \mathbb R$; \vspace{-0.25cm} 
\item[\hypertarget{(E)}{(E)}] 
Elliptic: a pair $(i \omega, -i\omega)$ for $\omega \in \mathbb R$; \vspace{-0.25cm}
\item[\hypertarget{(L)}{(L)}] 
Loxodromic: a quartet $(\gamma, -\gamma,\gamma^\ast,-\gamma^\ast)$ 
for $\gamma \in \mathbb C$. 
\end{itemize}
A Jordan decomposition can be employed through 
the construction of a transition matrix ${\bf Q}\in {\rm Mat}(2n,\mathbb R)$ 
such that ${\bf Q} \, \mathsf J \mathbf H \, {\bf Q}^{-1} = \mathbf J$, where $\bf J$ 
is the matrix composed by the Jordan blocks associated to the above eigenvalues. 
Consequently, the eigenvalues of $\mathsf S_t$ in (\ref{Eq:SympEv}) 
are exactly the exponential of the ones above. 
Actually, this classification holds for any 
$\mathsf S \in {\rm Sp}(2n,\mathbb R)$ \cite{arnold,gutzwiller}.          

Note that the elements of $\mathfrak{mp}(2n)$ 
and of $\mathfrak{sp}(2n)$ are in an one-to-one correspondence,   
which is not true for $\mathfrak{sp}(2n)$ and 
${\rm Sp}(2n,\mathbb R)$ since two distinct values of the parameter $t$
can give rise to the same symplectic matrix $\mathsf S_t$ in (\ref{Eq:SympEv}).   
This duplicity is controlled by a topological quantity associated 
to ${\rm Sp}(2n,\mathbb R)$ dubbed Conley-Zehnder index \cite{Conley1984}, 
which will be defined in the context of unitary operators, see Sec.\ref{mgczi}.   

Finally, the most generic \hy{QH} is 
\begin{equation}                                                                         \label{Eq:ClQuadLinHam}
H_{\rm cl} := H_1 + H_2 =  \tfrac{1}{2} x \cdot \mathbf H x + 
                           x \wedge \zeta  + H_0,
\end{equation}
which is a member of the inhomogeneous metaplectic Lie algebra denoted by 
${\text I}\mathfrak{mp}(2n)$ and generates the flux  
\begin{equation}                                                                         \label{Eq:FluxQuadLinHam}
\varphi(x_0,t) = \mathsf S_t x_0 + 
                 \int_0^t \!\! {\rm d}\tau \, {\mathsf S}_\tau \zeta \,\, , 
\,\,\,
\mathsf S_t = {\exp}[{\mathsf J {\bf H} t}].   
\end{equation} 
%
%%%%%%%%%%%%%%%%%%%%%%%%%%%%%%%%%%%%%%%%%%%%%%%%%%%%%%%%%%%%%%%%%%%%%%%%%%%%%%%%%%%%%%%%%
\subsection*{Wick Rotation of Classical Dynamics}
%%%%%%%%%%%%%%%%%%%%%%%%%%%%%%%%%%%%%%%%%%%%%%%%%%%%%%%%%%%%%%%%%%%%%%%%%%%%%%%%%%%%%%%%%
The action of the Wick rotation (\ref{Eq:WickRot}) on the classical flux is determined 
by the transformed Hamilton equations, {\it viz}, for a time independent Hamiltonian   
\begin{equation}                                                                         \label{Eq:ClasWick}
\frac{d x}{dt} \mapsto \frac{i}{\hbar} \frac{d x}{d\beta} = 
\mathsf J\frac{\partial H}{\partial x} 
\Longrightarrow 
\frac{d x}{d\beta} = - i \hbar \mathsf J \frac{\partial H}{\partial x}.  
\end{equation}
On this account, 
the flux generated by the \hy{QH} in (\ref{Eq:FluxQuadLinHam}) 
is 
\[
\varphi(x_0,\beta) = \mathsf S_{-i\hbar \beta} \, x_0  
                    -i \hbar \int_0^{-i\hbar\beta} \!\!\!\!\!\!\!\!\!\! 
                             {\rm d}\tau \, {\mathsf S}_\tau \zeta, 
\]
where the path of integration is along the imaginary axis and  
\begin{equation}                                                                         \label{Eq:CompSymp}
\mathtt S_{\beta} := 
\mathsf S_{-i\hbar \beta} = \exp[-i\hbar \beta \mathsf J \mathbf H] = 
(\mathtt S_{\beta}^\ast)^{-1} .                        
\end{equation}%

The Wick rotation, from (\ref{Eq:ClasWick}), is a simple multiplication of 
the generators by a complex constant:  
\begin{equation}                                                                         \label{Eq:WGen}
\mathsf J \mathbf H \mapsto -i \hbar \mathsf J \mathbf H , 
\,\,\,
\mathsf J \zeta \mapsto -i \hbar \mathsf J \zeta,
\end{equation}
which does not change the corresponding Lie algebras 
$\mathfrak{h}(2n)$ and $\mathfrak{sp}(2n)$ themselves. 
The matrix $\mathtt S_{\beta}$
belongs to the complex symplectic group 
${\rm Sp}(2n,\mathbb C) := \{\mathtt S \in {\rm Mat}(2n,\mathbb C) \, | \, 
\mathtt S^\top \mathsf J \mathtt S = \mathsf J\}$, 
since 
\[%\begin{equation}                                                                         \label{Eq:CompSymRel}
\begin{aligned}
\mathsf J \mathtt S_\beta^{\top} \mathsf J^\top 
= \mathsf J \exp[ i\hbar \beta \mathbf H \mathsf J ] \mathsf J^\top = 
\exp[ i\hbar \beta \mathsf J \mathbf H  ] = \mathtt S_\beta^{-1},   
\end{aligned}
\]%\end{equation}
which is equivalent to the symplectic condition and, 
accordingly to (\ref{Eq:sympPf}), 
$\det \mathtt S_{\beta} = 1$.  
Additionally, $\mathtt S_{\beta}$ satisfies 
\begin{equation}                                                                         \label{Eq:CompSympInv}
\mathtt S_{\beta}^\ast = \mathtt S_{\beta}^{-1}, \,\,\, 
{\rm Tr}\,{\mathtt S}_\beta \in \mathbb R,                       
\end{equation}
which are properties not shared by all matrices in ${\rm Sp}(2n,\mathbb C)$. 
It is thus convenient to define the Wick rotated version of the real symplectic group 
by 
\[
{\rm WSp}(2n,\mathbb C) := \{ \mathtt S \in {\rm Sp}(2n,\mathbb C) \,|\, 
                              \mathtt S^\ast = \mathtt S^{-1} \}
                               \subseteq {\rm Sp}(2n,\mathbb C).
\]

Due to (\ref{Eq:WGen}), 
${\rm WSp}(2n,\mathbb C)$ and ${\rm Sp}(2n,\mathbb R)$ share 
the same Lie algebra, {\it viz.}, $\mathfrak{sp}(2n)$ and  
the Wick rotation does preserve the categorization 
into \hy{(P)}, \hy{(H)}, \hy{(E)} and \hy{(L)}, since it is a constrain on 
the generators $\mathsf J\mathbf H \in \mathfrak{sp}(2n)$. 
However, 
the uniparametric subgroups of (\ref{Eq:SympEv}) and of (\ref{Eq:CompSymp}) 
can not belong always to the same category, 
due to the complex nature of the rotation:
categories \hy{(P)} and \hy{(L)} are invariant, 
while categories \hy{(H)} and \hy{(E)} are interchanged. 
In short,  
\begin{equation}                                                                         \label{Tab:WickClass}
t \mapsto -i\hbar \beta \Longrightarrow 
%\mathsf J \mathbf H \mapsto -i \hbar \mathsf J \mathbf H 
%\Longleftrightarrow
\left\{
\begin{array}{c}
\text{\hy{(P)}} \rightarrow \text{\hy{(P)}}; \\
\text{\hy{(L)}} \rightarrow \text{\hy{(L)}}; \\
\text{\hy{(H)}} \rightleftarrows \text{\hy{(E)}}.  
\end{array}
\right.
\end{equation}

By the end, 
the structural difference between 
the relation of the symplectic groups, real and complex, 
with theirs respective uniparametric subgroups should be highlighted, 
and relies on properties (\ref{Eq:CompSympInv}).
As already mentioned, a generic symplectic matrix ${\sf S} \in {\rm Sp}(2n,\mathbb R)$ 
has not the form in (\ref{Eq:SympEv}), 
however it is always decomposable as 
a product of matrices like the one in (\ref{Eq:SympEv}), 
of course, with different generators \cite{arvind,gossonbook2006}. 
In contrast, generic complex matrices in ${\rm Sp}(2n,\mathbb C)$ 
have complex trace, thus they can not be written as a matrix in (\ref{Eq:CompSymp}), 
neither can be decomposed as a product of such matrices.

% for generic matrices $\mathtt S_{1},\mathtt S_{2} \in {\rm Sp}(2n,\mathbb C)$, 
% the product $(\mathtt S_{1}\mathtt S_{2})^\ast \ne (\mathtt S_{1}\mathtt S_{2})^{-1}$.
%
% $\mathtt S'_{\beta'} := \exp[ -i\hbar \beta' \mathsf J \mathbf H']$ 
% is such that 
% ${\rm Tr}(\mathtt S_\beta \mathtt S'_{\beta'})^\dagger = 
% {\rm Tr}(\mathtt S_\beta^\dagger \mathtt S'^\dagger_{\beta'}) = 
% {\rm Tr}(\mathtt S_\beta \mathtt S'_{\beta'}) \in \mathbb R$;
% %
% thus not any matrix ${\tt S} \in {\rm Sp}(2n,\mathbb C)$ can be written as 
% products of uniparametric matrices like (\ref{Eq:CompSymp}).
% 
%%%%%%%%%%%%%%%%%%%%%%%%%%%%%%%%%%%%%%%%%%%%%%%%%%%%%%%%%%%%%%%%%%%%%%%%%%%%%%%%%%%%%%%%%
%%%%%%%%%%%%%%%%%%%%%%%%%%%%%%%%%%%%%%%%%%%%%%%%%%%%%%%%%%%%%%%%%%%%%%%%%%%%%%%%%%%%%%%%%
\section{Subgroups of Unitary Evolutions}                           \label{Sec:sue} %%%%%
%%%%%%%%%%%%%%%%%%%%%%%%%%%%%%%%%%%%%%%%%%%%%%%%%%%%%%%%%%%%%%%%%%%%%%%%%%%%%%%%%%%%%%%%%
%%%%%%%%%%%%%%%%%%%%%%%%%%%%%%%%%%%%%%%%%%%%%%%%%%%%%%%%%%%%%%%%%%%%%%%%%%%%%%%%%%%%%%%%%
Time-evolution operators of quantum systems constitutes a subgroup of the unitary 
operators. This section describes the unitary representation of those subgroups 
presented previously on Sec.\ref{Sec:se}. 
In addition, the group of translations and reflections will also be discussed, 
since it is necessary to perform calculations with the 
symbols in Sec.\ref{Sec:WWF}. 

%%%%%%%%%%%%%%%%%%%%%%%%%%%%%%%%%%%%%%%%%%%%%%%%%%%%%%%%%%%%%%%%%%%%%%%%%%%%%%%%%%%%%%%%%
\subsection{Heisenberg Group and Reflections}                        \label{SubSec:HGR}
%%%%%%%%%%%%%%%%%%%%%%%%%%%%%%%%%%%%%%%%%%%%%%%%%%%%%%%%%%%%%%%%%%%%%%%%%%%%%%%%%%%%%%%%%
Using the Zassenhaus (or BHC) formula \cite{QuantMech} and the \hy{CCR}, 
the composition of two Weyl operators in (\ref{Eq:HeisOp}) is 
\begin{equation}                                                                         \label{Eq:HeisComp}
\hat T_{\xi'} \hat T_{\xi''} = 
\exp\!\left[\frac{i}{2\hbar} {\xi'} \wedge \xi'' \right] \, 
\hat T_{\xi'+ \xi''};   
\end{equation}
thus the set of Weyl operators constitutes a continuous Lie group, 
which is a representation of the Heisenberg group $\text{H}(2n)$ 
\cite{gossonbook2006,littlejohn1986,yeh,ozorio1998}.     
The quantization of the classical Hamiltonian (\ref{Eq:ClLinHam}) with $H_0 = 0$,  
$H_1(\hat x) = \hat x \wedge \zeta$, 
is the generator of an uniparametric subgroup of Weyl operators. 
Like its classical version, 
the set of these Hamiltonians constitutes the Lie algebra $\mathfrak{h}(2n)$ 
under the commutator operation.   

In contrast, a reflection, as parity, is not continuous 
and their set even constitutes a group, both of which are evident from   
the composition rule  
\begin{equation}                                                                         \label{Eq:RefComp}
{\hat R}_{x'}{\hat R}_{x''} = 
\exp\!\left[-\frac{2i}{\hbar}  x' \wedge x'' \right] \hat T_{2(x'-x'')},
\end{equation}
that can be derived from (\ref{Eq:ReflOp}). 
However, following \cite{ozorio1998}, 
the composition of a reflection and a translation, 
\begin{equation}                                                                         \label{Eq:RefTrComp}
{\hat R}_{x} \hat T_{\xi} = \exp\!\left[\frac{i}{\hbar} {\xi} \wedge x \right]
\hat R_{ x - \frac{\xi}{2} },
 \end{equation}
is a reflection. 
The product rules (\ref{Eq:ReflOp}), (\ref{Eq:RefComp}) and (\ref{Eq:RefTrComp}) 
show that the set ${\rm Oz}(2n):= {\rm H}(2n) \cup \{\hat R_0\}$ 
is a (discrete) group of translations and reflections \cite{ozorio1998}. 

From the definitions in (\ref{Eq:WWsymbs}), 
using accordingly the compositions (\ref{Eq:RefComp}) or (\ref{Eq:RefTrComp}), 
and Eq.(\ref{Eq:TRtrace}), the symbols of the elements in ${\rm Oz}(2n)$ are 
\begin{equation}                                                                         \label{Eq:TransRefSymb}
\begin{aligned}
& \tilde T_\eta(\xi) = (2\pi\hbar)^n \delta^{2n} (\eta -\xi), \,\,\,
T_\eta(x) = {\rm e}^{-\frac{i}{\hbar} x \wedge \eta};                           \\
& \tilde R_x(\xi) = \frac{{\rm e}^{\frac{i}{\hbar} x \wedge \xi}}{2^n} , \,\,\,
R_\eta(x) = (\pi\hbar)^n \delta^{2n} (x-\eta).  
\end{aligned}
\end{equation}

%%%%%%%%%%%%%%%%%%%%%%%%%%%%%%%%%%%%%%%%%%%%%%%%%%%%%%%%%%%%%%%%%%%%%%%%%%%%%%%%%%%%%%%%%
\subsection{Metaplectic Group and Conley-Zehnder Index}\label{mgczi}
%%%%%%%%%%%%%%%%%%%%%%%%%%%%%%%%%%%%%%%%%%%%%%%%%%%%%%%%%%%%%%%%%%%%%%%%%%%%%%%%%%%%%%%%%
For $\mathsf S \in {\rm Sp}(2n,\mathbb R)$, 
an unitary operator $\hat M_\mathsf{S}$ such that  
\begin{equation}                                                                         \label{Eq:MetOpDef}
\hat M_\mathsf{S}^\dagger \hat x \hat M_\mathsf{S} = \mathsf S \hat x 
\end{equation}
is called a Metaplectic operator (\hypertarget{MO}{MO}) and is a member of the 
subgroup of the unitary operators called Metaplectic group 
and denoted by ${\rm Mp}(2n)$ \cite{gossonbook2006,littlejohn1986}.

The symmetric quantization of (\ref{Eq:ClQuadHam}),  
\begin{equation}                                                                         \label{Eq:QuQuadHam}
\hat H_2 = \frac{1}{2}\hat x \cdot {\bf H} \hat x,  
\end{equation}
is an element of the metaplectic Lie algebra $\mathfrak{mp}(2n)$, 
the same algebra of the classical Hamiltonians (\ref{Eq:ClQuadHam}), 
but now under the commutator operation \cite{gossonbook2006,littlejohn1986}.  
Thus, an uniparametric subgroup of ${\rm Mp}(2n)$ is constituted by  
\begin{equation}                                                                         \label{Eq:MetapEv}
\hat M_{\mathsf S_{t}} := 
{\exp}\left[{-\frac{i t}{2\hbar} \hat x \cdot {\bf H} \hat x}\right],   
\end{equation}
where the subindex $\mathsf S_t$ highlights the relation between 
the \hy{MO} and the symplectic matrix (\ref{Eq:SympEv}). 
It is important to stress that not all operators defined through (\ref{Eq:MetOpDef}) 
are like the ones in (\ref{Eq:MetapEv}).  

As for the classical dynamics, 
for each element (\ref{Eq:QuQuadHam}) in $\mathfrak{mp}(2n)$, 
which is in one-to-one correspondence with $\mathfrak{sp}(2n)$, 
there are two in ${\rm Mp}(2n)$ and it is said that the Metaplectic group 
is a double covering group of the symplectic one \cite{gossonbook2006}. 
This will be clear from the analyses of the symbols related to 
the \hy{MO}s.

The Weyl and Wigner symbols (\ref{Eq:WWsymbs}) of a generic \hy{MO} 
are given \cite{mehlig,ozorio1998,gossonbook2006}, respectively, by 
\begin{equation}                                                                         \label{Eq:Hsmet}
\tilde{M}_\mathsf{S}(\xi) = 
\frac{
\exp\left[
-\frac{i}{4\hbar} \xi \cdot 
\mathsf{J}\mathbf{C}_{\mathsf S}^{-1}
\mathsf{J}\xi\right]}
{\sqrt{\det\left(\mathsf S - \mathsf I_{2n}\right)}},                                                          
\end{equation}
and
\begin{equation}                                                                         \label{Eq:Wsmet}   
M_\mathsf{S}(x) =  
\frac{2^n 
\exp\left[-\frac{i}{\hbar} x \cdot \mathbf{C}_{\mathsf S}x\right]}
{\sqrt{\det\left(\mathsf S + \mathsf I_{2n}\right)}},           
\end{equation}
where ${\bf C}_\mathsf{S} \in {\rm Mat}(2n,\mathbb R)$ 
stands for the {\it Cayley parametrization} of 
$\mathsf S \in {\rm Sp}(2n,\mathbb R)$ defined by 
\begin{equation}                                                                         \label{Eq:Cayley}
{\bf C}_\mathsf{S} := 
-\mathsf{J} \frac{\left(\mathsf{S} - \mathsf{I}_{2n} \right)} 
{\left(\mathsf{S} + \mathsf{I}_{2n}\right)} 
= {\bf C}_\mathsf{S}^\top = - {\bf C}_\mathsf{S^{-1}}.  
\end{equation}

Accordingly with $\mathsf S$, 
one or both of the above symbols may not be defined,
which happens when $\pm 1 \in {\rm Spec}(\mathsf S)$, 
{\it i.e.}, when 
$\det\left( \mathsf S \pm \mathsf I_{2n} \right)= 0$. 
These discontinuities are definitively not a property of the operator itself,    
it is an unavoidable feature of the expansions (\ref{Eq:WWreps}) for this class of operators.   
However, there are two available expansions, 
the divergences can be overcomed switching between the Weyl and Wigner representations. 
% However, this brings up a continuity question about the \hy{MO}:
% since there are two \hy{MO}s 
% [each one associated to a sign of the denominator in (\ref{Eq:Hsmet})
% or equivalently in (\ref{Eq:Wsmet})] 
% for one $\mathsf S_t$, it is necessary to keep track of the correct sign.  

The {\it Conley-Zehnder} (\hypertarget{CZ}{CZ}) index \cite{Conley1984} 
is an integer function 
$\nu^{-}_{\mathsf Q}: t \in \mathbb R_\ge \longmapsto \{0,1,2,3\}$, 
whose vocation is to count how many times a path 
$t \mapsto {\mathsf Q}_t \in {\rm Sp}(2n,\mathbb R)$ 
crosses the manifold $\det ({\mathsf Q}_t -\mathsf I_{2n}) = 0$ 
and in which direction the crossing occurs, {\it i.e.}, 
from negative to positive values or from positive to negative. 
This index and its companion $\nu^{+}_{\mathsf Q}$ are defined through 
\begin{equation}                                                                         \label{Eq:CZdef}
\sqrt{\det\left({\mathsf Q}_{t} \pm \mathsf I_{2n}\right)} =  
i^{ - \nu^{\pm}_{{\mathsf Q}_{t}}} 
\sqrt{\left |\det\left( {\mathsf Q}_{t} \pm \mathsf I_{2n} \right) \right|}   
\end{equation}
and both acquire the values in $\{0,2\}$ if 
$\det\left({\mathsf Q}_{t} \pm \mathsf I_{2n}\right)>0$, 
or in $\{1,3\}$ if $\det\left({\mathsf Q}_{t} \pm \mathsf I_{2n}\right)<0$. 
Note that the denominators in  
(\ref{Eq:Hsmet}) and in (\ref{Eq:Wsmet}) can be rewritten as (\ref{Eq:CZdef}).
%
% \[%\begin{equation}                                                                         \label{Eq:IndexDef}
% \sqrt{\det\left(\mathsf S \pm \mathsf I_{2n}\right)} =  i^{ - \nu^{\pm}_{\mathsf S}} 
% \sqrt{\left |\det\left( \mathsf S \pm \mathsf I_{2n} \right) \right|}   
% \]%\end{equation}
% and note that, for the case of the \hy{CZ}-index, 
% this is equivalent to the definition (\ref{Eq:CZdef}).  
%

If ${\tilde M}_\mathsf{S}(\xi)$ and $M_\mathsf{S}(x)$ are both well behaved, 
they are related by (\ref{Eq:SympFour}) and, for a given $\nu^{-}_{\mathsf S}$, 
the index $\nu^{+}_{\mathsf S}$ becomes
\begin{equation}                                                                         \label{Eq:FourierCZ}
\nu^{+}_{\mathsf S} = \nu^{-}_{\mathsf S} + 
                      \tfrac{1}{2} {\rm Sng}\,  \mathbf{C}_{\mathsf S} 
                      \, ({\rm mod} \, 4), 
\end{equation}
where ${\rm Sng} \, \bf X$ is the number of positive eigenvalues minus 
the number of negative eigenvalues of the matrix $\bf X$.  
Since both symbols does not diverge, 
then $\det\left( \mathsf S \pm \mathsf I_{2n} \right) \ne 0$ and 
${\rm Rank} \mathbf{C}_{\mathsf S} = 2n$, 
thus $\tfrac{1}{2} {\rm Sng}\,  \mathbf{C}_{\mathsf S} \in \mathbb Z$ 
and $\nu^{+}_{\mathsf S} \in \{0,1,2,3\}$.
The Fourier relation (\ref{Eq:SympFour}) for the symbols in question should 
be faced as Fresnel-type integral \cite{gossonbook2006}. 

From the above discussion, 
the main features of the \hy{CZ}-index become clear:  
each of the two \hy{MO}s (double coverage) associated to 
one symplectic matrix is distinguished by one value of the \hy{CZ}-index, 
which in turn is associated to the sign of the square-root which appears 
in (\ref{Eq:Hsmet}) or in (\ref{Eq:Wsmet}); 
rather than mere signs, these indexes are  
imposed by the continuity of the operators when 
changing between representations by (\ref{Eq:FourierCZ}), 
or when a symbol crosses a divergence. 

The \hy{CZ}-index and some aspects of the \hy{MO}s have been dealing 
here on a broader scope. 
However, in this work the treatment will focus only on the relationship between 
the uniparametric subgroups of ${\rm Sp}(2n,\mathbb R)$ and of ${\rm Mp}(2n)$ 
composed, respectively, 
by matrices of the form (\ref{Eq:SympEv}) and operators as in (\ref{Eq:MetapEv})% 
%
%%%%%%%%%%%%%%%%%%%%%%%%%%%%%%%%%%%%%%%%%%%
\footnote{%   
The complete representation theory between 
${\rm Sp}(2n,\mathbb R)$ and ${\rm Mp}(2n)$ 
is developed in \cite{gossonbook2006}.}. 
%%%%%%%%%%%%%%%%%%%%%%%%%%%%%%%%%%%%%%%%%%%
%
The protocol for the Weyl-Wigner symbols of \hy{MO}s in (\ref{Eq:MetapEv}) 
which describes when symbols (\ref{Eq:Hsmet}) and (\ref{Eq:Wsmet}) 
attains a divergence will be briefly mentioned, 
inasmuch as all details and examples can be found in \cite{nicacio12}. 
 
The continuity of a \hy{MO} in (\ref{Eq:MetapEv}) demands  
\begin{equation}                                                                         \label{Eq:CZlimit}
\lim_{t \to 0^+} \mathsf S_t = {\sf I}_{2n} \Longrightarrow
\lim_{t \to 0^+} \hat M_{\mathsf{S}_{t}}  =  + \hat{\sf 1}.
\end{equation}
Thus, for $t = 0$,  
the Weyl symbol $\tilde{M}_{\mathsf I_{2n}} (\xi)$ is not defined. 
However, the Wigner symbol of the identity operator should be  
$M_{\mathsf I_{2n}} (x) = + 1$ from (\ref{Eq:CZlimit}), 
which shows that $\nu^{+}_{\mathsf I_{2n}} = 0$.
The imposition of the initial condition (\ref{Eq:CZlimit}) 
restricts the values of the indexes throughout 
the whole evolution dictated by $\mathsf S_{t}$.    
Supposing that the two symbols (\ref{Eq:Hsmet}) and (\ref{Eq:Wsmet}) 
never diverge at the same instant, 
just before a divergence of one symbol, 
relation (\ref{Eq:SympFour}), or its inverse, should be used to   
determine the other. Thus after the same divergence,
the original symbol is recovered also using the inverse, or the direct version, 
of (\ref{Eq:SympFour}). 
For each Fourier transformation, 
the appropriate CZ-index is determined by (\ref{Eq:FourierCZ}), 
and note that, out of a divergence, 
${\rm Sng}\,  \mathbf{C}_{\mathsf S_t} = 
{\rm Sng}\,  \mathbf{C}_{\mathsf S_t}^{-1}$.

\subsection{Inhomogeneous Metaplectic Group}
%%%%%%%%%%%%%%%%%%%%%%%%%%%%%%%%%%%%%%%%%%%%%%%%%%%%%%%%%%%%%%%%%%%%%%%%%%%%%%%%%%%%%%%%%
The composition of a generic \hy{MO} with the Heisenberg translation (\ref{Eq:HeisOp}) 
%and with a Reflection operator (\ref{Eq:ReflOp}) are 
is a direct consequence of (\ref{Eq:MetOpDef}):
\begin{equation}                                                                         \label{Eq:CovRel}
\hat M_{\mathsf S} \hat T_\xi \hat M_{\mathsf S}^\dagger = 
\hat T_{{\mathsf S}\xi}.  
%\,\,\,
%\hat M_{\mathsf S} \hat R_x \hat M_{\mathsf S}^\dagger = \hat R_{{\mathsf S} x}.  
\end{equation}
This relation turns to be possible the definition of 
the Inhomogeneous Metaplectic group 
${\rm IMp}(2n)$ which is composed by operators of the form   
\begin{equation}                                                                         \label{Eq:IMop}
\hat U = {\rm e}^{\frac{i}{\hbar} \phi} \hat T_\xi \hat M_{\mathsf S} ,
\end{equation}
where $\hat T_\xi \in {\rm H}(2n)$, 
$\hat M_{\mathsf S} \in {\rm Mp}(2n)$, and 
$ \phi \in \mathbb R$.
%, such that ${\rm e}^{\frac{i}{\hbar} \varphi} \in {\rm U}(1)$.   

An uniparametric subgroup of ${\rm IMp}(2n)$ has it elements conveniently written as  
\begin{equation}                                                                          \label{Eq:UnitQuadLinHam}
\hat U_t =  
            \hat T_{\eta} \, \hat M_{{\sf S}_t}\hat T_{\eta}^\dagger \,  
            {\rm e}^{-\tfrac{it}{\hbar} H_0}, 
\end{equation}
with $\hat T_{\eta}$ in (\ref{Eq:HeisOp}), 
and $\hat M_{\mathsf S_t}$ in (\ref{Eq:MetapEv}). 
The above unitary operator is indeed a member of ${\rm IMp}(2n)$, 
since $\hat U_t$ assumes the form in (\ref{Eq:IMop}) 
when relations (\ref{Eq:CovRel}) and (\ref{Eq:HeisComp}) are applied. 
Generators of this subgroup are determined through  
\begin{equation}                                                                         \label{Eq:QQuadLinham}
\hat H := i\hbar \hat U_t^{-1}\frac{d\hat U_t}{dt}  = 
\frac{1}{2} \hat x \cdot \mathbf H \hat x + 
\hat x \wedge \zeta  + H'_0,   
\end{equation}
which is the symmetric quantization of the Hamiltonian (\ref{Eq:ClQuadLinHam}) 
with $\zeta := - \mathsf J \mathbf H \eta$ and 
$H'_0 = H_0 + \tfrac{1}{2}\eta {\bf H} \eta$,   
thus a member of ${\rm I}\mathfrak{mp}(2n)$ 
Lie algebra \cite{gossonbook2006,littlejohn1986}.  

The Weyl symbol in (\ref{Eq:UnitOpSymb}) 
of the unitary operator in (\ref{Eq:UnitQuadLinHam}) 
is obtained using the compositions rules of translations 
and reflections in Sec.\ref{SubSec:HGR}, 
the symplectic covariance relation (\ref{Eq:CovRel}), 
and the ciclicity of the trace. Indeed, 
\begin{eqnarray}                                                                         \label{Eq:WeylGenQuad}
\tilde U_t (\xi) &=&  
                    {\rm Tr}\left(  
                                  \hat M_{{\sf S}_t} \,  
                                  \hat T_{\eta}^\dagger 
                                  \hat T_\xi^\dagger 
                                  \hat T_{\eta}  \right)
                    {\rm e}^{- \tfrac{i}{\hbar} H_0 t }                                \nonumber \\
                &=& {\rm Tr}\left(  
                                    \hat M_{{\sf S}_t} \,  
                                    \hat T_{\xi}^\dagger \right)
                     {\rm e}^{   
                               - \tfrac{i}{\hbar} H_0 t 
                               - \tfrac{i}{\hbar}\xi\wedge\eta }                       \\
                &=&  \tilde M_{{\sf S}_t}(\xi) \,\,              
                     {\rm e}^{   
                               - \tfrac{i}{\hbar} (H_0 t + \xi\wedge\eta) },             \nonumber
\end{eqnarray}
where $\tilde M_{{\sf S}_t}(\xi)$ is defined by (\ref{Eq:Hsmet}).
Similarly, the Wigner symbol becomes
\begin{eqnarray}                                                                         \label{Eq:WignerlGenQuad}
U_t (x) &=& M_{{\sf S}_t}(x-\eta) \, 
{\rm e}^{ - \tfrac{i}{\hbar}  H_0 t },
\end{eqnarray}
where $ M_{{\sf S}_t}(x)$ is defined by (\ref{Eq:Wsmet}).
%%%%%%%%%%%%%%%%%%%%%%%%%%%%%%%%%%%%%%%%%%%%%%%%%%%%%%%%%%%%%%%%%%%%%%%%%%%%%%%%%%%%%%%%%
\subsection{Covariances of Operators and Symbols}                         \label{Sec:cos}
%%%%%%%%%%%%%%%%%%%%%%%%%%%%%%%%%%%%%%%%%%%%%%%%%%%%%%%%%%%%%%%%%%%%%%%%%%%%%%%%%%%%%%%%%
The covariance relation (\ref{Eq:CovRel}) is the one of, if not, 
the major advantage in working with a basis of operators which sets
momentum and position on equal footing. 
Consider $\mathsf Q \in {\rm Sp}(2n,\mathbb R)$, $\zeta \in \mathbb R^{2n}$, 
and the operator defined as 
\begin{equation}                                                                         \label{Eq:QuadTransf}
\hat A' := ( \hat M_{\mathsf Q} \hat T_{\zeta})^\dagger 
\hat A 
(\hat M_{\mathsf Q}\hat T_{\zeta} ).
\end{equation}
Taking the Weyl-Wigner expansion of $\hat A$ in (\ref{Eq:WWreps}), 
by virtue of the composition formulas (\ref{Eq:HeisComp}), (\ref{Eq:RefTrComp}) 
and of (\ref{Eq:CovRel}), the symbols are covariant with $\hat A$: 
\begin{equation}                                                                         \label{Eq:SymbsCov}
\tilde A'(\xi) := A({\mathsf Q} \xi) \, {\rm e}^{\frac{i}{\hbar} \xi \wedge \zeta}, 
\,\,\,\, 
A'(x) := A({\mathsf Q}(x + \zeta)). 
\end{equation}

Applying the above covariance rule 
for the symbols of the metaplectic operator (\ref{Eq:Hsmet}) and (\ref{Eq:Wsmet}), 
one finds 
\[                                                                                          
\begin{aligned}
\tilde M'_\mathsf{S}(\xi) & = 
\tilde M_\mathsf{S}({\mathsf Q} \xi ) \, {\rm e}^{\frac{i}{\hbar} \xi \wedge \zeta} = 
\tilde M_\mathsf{Q^{-1} S Q}( \xi ) \, {\rm e}^{\frac{i}{\hbar} \xi \wedge \zeta},      \\
M'_\mathsf{S}(x) & = 
M_\mathsf{S}({\mathsf Q}( x + \zeta)) = 
M_\mathsf{Q^{-1} S Q}( x + \zeta),    
\end{aligned}
\]
where the rightmost equalities in both equations are consequences of
the covariance of the Cayley transform in (\ref{Eq:Cayley}),  
\begin{equation}                                                                         \label{Eq:CayleyCov}
{\mathsf Q}^{\top} {\bf C}_\mathsf{S} {\mathsf Q} = {\bf C}_\mathsf{Q^{-1} S Q}, 
\end{equation}
and of $\det\left(\mathsf{Q^{-1} S Q} \pm \mathsf I_{2n} \right) = 
\det\left( \mathsf S \pm \mathsf I_{2n} \right)$. 
Since ${\mathsf Q}^{\top} {\bf C}_\mathsf{S} {\mathsf Q}$ is a 
congruence of ${\bf C}_\mathsf{S}$, thus 
\begin{equation}                                                                         \label{Eq:SignCayleyCov}
{\rm Sng}\,  \mathbf{C}_{\mathsf S} = 
{\rm Sng} ({\mathsf Q}^{\top} {\bf C}_\mathsf{S} {\mathsf Q}) =
{\rm Sng} {\bf C}_\mathsf{Q^{-1} S Q}. 
\end{equation}
The last invariance relations, by (\ref{Eq:CZdef}) and (\ref{Eq:FourierCZ}), 
also guarantee the invariance of the \hy{CZ} index:   
\[%\begin{equation}                                                                         \label{Eq:CZinv}
\nu^{\pm}_{\mathsf{Q^{-1} S Q}} (t) = \nu^{\pm}_{\mathsf S} (t).
\]%\end{equation}
By the end, since $\mathsf{Q^{-1} S Q}$ is similar to $\mathsf S$, 
its eigenvalues are invariant. 

If the evolution is governed by Hamiltonian (\ref{Eq:QQuadLinham}), 
the unitary operator in (\ref{Eq:UnitQuadLinHam}), 
after the transformation in (\ref{Eq:QuadTransf}), becomes   
\[
\hat U_t' = 
\hat T_{{\mathsf Q}^{-1}\eta -\zeta}\, 
\hat M_{{\mathsf Q}^{-1} \mathsf{S}_{t} {\mathsf Q}}\,  
\hat T_{{\mathsf Q}^{-1}\eta -\zeta}^\dagger \, {\rm e}^{-\frac{i}{\hbar} H_0 t}.
\] 
The intrinsic relation of
$\hat M_{{\mathsf Q}^{-1} \mathsf{S}_{t} {\mathsf Q}}$ in above formula 
with the metaplectic evolution in (\ref{Eq:MetapEv}) is reveled 
by Eq.(\ref{Eq:MetOpDef}):    
\[
\hat M_{{\mathsf Q}^{-1} \mathsf{S}_{t} {\mathsf Q}} = 
\hat M_{{\mathsf Q}}^\dagger \hat M_{\mathsf{S}_{t}} \hat M_{{\mathsf Q}} = 
{\exp}\left[{-\frac{i t}{2\hbar} \hat x \cdot 
                                 {\mathsf Q}^\top{\bf H} {\mathsf Q} \hat x}\right],    
\]   
which is the metaplectic operator associated to covariant symplectic matrix 
\[
\mathsf S'_{t} = 
\exp[\mathsf J {\mathsf Q}^\top{\bf H} {\mathsf Q} t] = 
\exp[{\mathsf Q}^{-1} \mathsf J {\bf H} {\mathsf Q} t] = 
{\mathsf Q}^{-1} \mathsf S_{t} {\mathsf Q}. 
\]
Since ${\bf H}' = {\mathsf Q}^\top{\bf H} {\mathsf Q}$ and 
$ \mathsf J{\bf H}' = {\mathsf Q}^{-1} \mathsf J {\bf H} {\mathsf Q}$,
then ${\rm Sng}{\bf H}' = {\rm Sng}{\bf H}$ and the spectrum of 
$\mathsf J \mathbf H$ are invariant, 
as well as the spectrum of $\mathsf S_t$. 

%%%%%%%%%%%%%%%%%%%%%%%%%%%%%%%%%%%%%%%%%%%%%%%%%%%%%%%%%%%%%%%%%%%%%%%%%%%%%%%%%%%%%%%%%
%%%%%%%%%%%%%%%%%%%%%%%%%%%%%%%%%%%%%%%%%%%%%%%%%%%%%%%%%%%%%%%%%%%%%%%%%%%%%%%%%%%%%%%%%
\section{Thermal Operators of Quadratic Hamiltonians}                    \label{Sec:toqh}
%%%%%%%%%%%%%%%%%%%%%%%%%%%%%%%%%%%%%%%%%%%%%%%%%%%%%%%%%%%%%%%%%%%%%%%%%%%%%%%%%%%%%%%%%
%%%%%%%%%%%%%%%%%%%%%%%%%%%%%%%%%%%%%%%%%%%%%%%%%%%%%%%%%%%%%%%%%%%%%%%%%%%%%%%%%%%%%%%%%
The Weyl-Wigner symbols of the thermal operator and the \hy{PF}
for a general \hy{QH} will be calculated through the mathematical tools 
presented in all previous sections. Thermodynamical properties and the classical limit 
for the \hy{QH}s will also be developed.  
%%%%%%%%%%%%%%%%%%%%%%%%%%%%%%%%%%%%%%%%%%%%%%%%%%%%%%%%%%%%%%%%%%%%%%%%%%%%%%%%%%%%%%%%%
\subsection{Weyl-Wigner Representation}
%%%%%%%%%%%%%%%%%%%%%%%%%%%%%%%%%%%%%%%%%%%%%%%%%%%%%%%%%%%%%%%%%%%%%%%%%%%%%%%%%%%%%%%%%
The Cayley parametrization (\ref{Eq:Cayley}) of the complex symplectic matrix 
in (\ref{Eq:CompSymp}) is an anti-Hermitian matrix, 
since it is symmetric and 
\begin{equation}                                                                         \label{Eq:CompCayley}
{\rm Re} \mathbf{C}_{{\mathtt S}_{\beta}} = 0 \Longleftrightarrow
\mathbf{C}_{{\mathtt S}_{\beta}}^\ast = 
\mathbf{C}_{{\mathtt S}_{\beta}^\ast} =
\mathbf{C}_{{\mathtt S}_{\beta}^{-1}} = 
- \mathbf{C}_{{\mathtt S}_{\beta}},      
 \end{equation}
due to the property in (\ref{Eq:CompSympInv}).
The formula for $\mathbf{C}_{{\mathtt S}_{\beta}}$ 
can be manipulated to write it  
in terms of the real and complex components of 
$\mathtt S_{\beta}$. 
Indeed, if $-1 \notin {\rm Spec}({\mathtt S}_{\beta})$, 
from (\ref{Eq:Cayley}) and using (\ref{Eq:CompSympInv}), 
\[
\begin{aligned}
\mathbf{C}_{{\mathtt S}_{\beta}} &=  
\mathsf J^\top 
({\mathtt S}_{\beta} + \mathsf I_{2n})^{-1} 
({\mathtt S}_{\beta}^{\ast} + \mathsf I_{2n})^{-1} 
({\mathtt S}_{\beta}^{\ast} + \mathsf I_{2n})
({\mathtt S}_{\beta} - \mathsf I_{2n})       \\
& = i \mathsf J^\top 
({\rm Re}{\mathtt S}_{\beta} + \mathsf I_{2n})^{-1} 
{\rm Im}{\mathtt S}_{\beta} . % \\
%
% & \Longleftrightarrow 
% {\rm Im}\mathbf{C}_{{\mathtt S}_{\beta}} =  
% \mathsf J^\top 
% ({\rm Re}{\mathtt S}_{\beta} + \mathsf I_{2n})^{-1} 
% {\rm Im}{\mathtt S}_{\beta}.                         
\end{aligned}
\]
For the case in which $1 \notin {\rm Spec}({\mathtt S}_{\beta})$, 
a similar procedure gives
\[
\begin{aligned}
\mathbf{C}_{{\mathtt S}_{\beta}}^{-1} &= 
i({\rm Re}{\mathtt S}_{\beta} - \mathsf I_{2n})^{-1}
({\rm Im}{\mathtt S}_{\beta})
\mathsf J^\top. %                      \\
% & \Longleftrightarrow 
% ({\rm Im}\mathbf{C}_{{\mathtt S}_{\beta}})^{-1} =  
% ({\rm Re}{\mathtt S}_{\beta} - \mathsf I_{2n})^{-1}
% ({\rm Im}\mathbf{C}_{{\mathtt S}_{\beta}})
% \mathsf J.                        
\end{aligned}
\]
Note also that 
\begin{equation}                                                                         \label{Eq:Prop}
\!\!\!\!\begin{aligned} 
&\det\mathbf{C}_{{\mathtt S}_{\beta}} = 
(-1)^n \det {\rm Im} \mathbf{C}_{{\mathtt S}_{\beta}} \in \mathbb R,  \\
&\det ({{\mathtt S}_{\beta}^\ast} \pm \mathsf I_{2n}) \! = 
\det ({{\mathtt S}_{\beta}^{-1}} \!\pm \mathsf I_{2n}) =
\det ({{\mathtt S}_{\beta}} \pm \mathsf I_{2n}) \in \mathbb R,  
\end{aligned}
\end{equation}
and remember that the action of the Wick rotation preserves the categorization 
of the eigenvalues, see Eq.(\ref{Tab:WickClass}).

The Weyl symbol of the thermal operator 
$\exp[{-\beta \hat H}]$ for the Hamiltonian (\ref{Eq:QQuadLinham}) 
is readily determined by the mapping (\ref{Eq:WickThSymb}) of 
the symbol (\ref{Eq:WeylGenQuad}),   
\begin{equation}                                                                         \label{Eq:WeylSymThermal}
\tilde{E}_\beta(\xi) =  
\frac{ 
      {\rm e}^{  
               -\tfrac{1}{4\hbar} 
                \xi \cdot \mathsf{J}
                          ({\rm Im} \mathbf{C}_{{\mathtt S}_{\beta}})^{-1}
                          \mathsf{J} \xi 
               -\beta H_0 } }
     { \sqrt{ \det( \mathtt S_{\beta}  - \mathsf I_{2n} ) }                   } 
 {\rm e}^{-\frac{i}{\hbar} \xi \wedge \eta} ,  
\end{equation}
and of the Wigner symbol (\ref{Eq:WignerlGenQuad}),                                             
\begin{equation}                                                                         \label{Eq:WigSymThermal}
{E}_\beta(x)   =    
\frac{ 2^n \, 
       {\rm e}^{\tfrac{1}{\hbar} (x-\eta) \cdot 
                                  {\rm Im} \mathbf{C}_{{\mathtt S}_{\beta}} 
                                  (x-\eta) - \beta H_0} }
     { \sqrt{ \det ( {\mathtt S}_{\beta} + \mathsf I_{2n} )} }.
\end{equation} 

As before, the Fourier transformation links above symbols when both are well behaved, 
{\it i.e.}, when $\pm 1 \notin {\rm Spec}(\mathtt S_{\beta})$. 
However, the convergence of (\ref{Eq:SympFour}) now is more restrictive, 
since ${\rm Im} \mathbf{C}_{{\mathtt S}_{\beta}} \in {\rm Mat}(2n,\mathbb R)$.  
It will be only guarantied if the integrand in (\ref{Eq:SympFour}) 
is an absolutely integrable function \cite{bracewell}. 
For the symbols in question, this means  
${\rm Sng} \, {\rm Im} \mathbf{C}_{{\mathtt S}_{\beta}} = - 2n$, 
that is, if ${\rm Im} \mathbf{C}_{{\mathtt S}_{\beta}} < 0$. 
This further requirement is a consequence of the mentioned matrix being real, 
and thus the integration does not rely any more on a Fresnel prescription, 
as it was before.  
Furthermore,
the indexes defined through (\ref{Eq:CZdef}) 
are mapped naturally by (\ref{Eq:WickRot}) to 
\begin{equation}                                                                         \label{Eq:WickCZ}
\sqrt{\det\left({\mathtt S}_{\beta} \pm \mathsf I_{2n}\right)} =  
i^{ - \nu^{\pm}_{{\mathtt S}_{\beta}}} 
\sqrt{\left |\det\left( {\mathtt S}_{\beta} \pm \mathsf I_{2n} \right) \right|},   
\end{equation}
also subjected to a physical ``initial condition'': 
\begin{equation}                                                                         \label{Eq:WCZlimit}
\lim_{\beta \to 0^+} {\mathtt S}_{\beta} = {\sf I}_{2n} \Longrightarrow
\lim_{\beta \to 0^+} {\rm e}^{-\beta \hat H}  =  + \hat{\sf 1}.
\end{equation}
Considering thus that all the requirements are met, 
the Fourier transformation gives
\begin{equation}                                                                         \label{Eq:WickCZ2}
\nu^{-}_{{\mathtt S}_{\beta}} = \nu^{+}_{{\mathtt S}_{\beta}} . 
\end{equation}

From Eqs.(\ref{Eq:ThWigChFunc}),   
the characteristic and Wigner functions of a thermal state of 
a \hy{QH} are, respectively,  
\begin{equation}                                                                         \label{Eq:WigCharFunc}
\begin{aligned}
\chi(\xi) &= 
\frac{{\rm e}^{  
               -\tfrac{1}{4\hbar} 
                \xi \cdot \mathsf{J}
                          ({\rm Im} \mathbf{C}_{{\mathtt S}_{\beta}})^{-1}
                          \mathsf{J} \xi -\frac{i}{\hbar} \xi \wedge \eta }}{(2\pi\hbar)^n} , \\ 
W(x) &= \frac{ 
              {\rm e}^{ \tfrac{1}{\hbar} 
                        (x-\eta) \cdot 
                        {\rm Im} \mathbf{C}_{{\mathtt S}_{\beta}} 
                        (x-\eta) } }
            { (\pi\hbar)^n 
              \sqrt{ \det {\rm Im} \mathbf{C}_{{\mathtt S}_{\beta}}^{-1}}}.
\end{aligned}
\end{equation}
If ${\rm Im} \mathbf{C}_{{\mathtt S}_{\beta}} < 0$,  
the Wigner function will be a normalized Gaussian with covariance matrix 
$\tfrac{\hbar}{2}{\rm Im} \mathbf{C}_{{\mathtt S}_{\beta}}^{-1}$ 
and mean-value $\eta$, this condition is satisfyed 
only by Hamiltonians in category \hy{(E)}, see Sec.\ref{Sec:EH}.
Otherwise, the Wigner function is not defined, 
since it will not provide the correct probability marginals 
of the quantum state \cite{ozorio1998}.   
However, 
the thermal operator can still be represented through its 
symbols in (\ref{Eq:WeylSymThermal}) and (\ref{Eq:WigSymThermal}), 
or even by its characteristic function, 
since none of these are subjected to the expected properties of a 
genuine Wigner Function. 

The Cayley transform (\ref{Eq:Cayley}) can be uniquely inverted, 
if $-1 \notin {\rm Spec}({\mathtt S}_{\beta})$, to give
\[
{\mathtt S}_{\beta} = 
(\mathsf I_{2n}+\mathsf J \mathbf{C}_{{\mathtt S}_{\beta}})
(\mathsf I_{2n}-\mathsf J \mathbf{C}_{{\mathtt S}_{\beta}})^{-1},
\]
and a similar relation can be written when 
$1 \notin {\rm Spec}({\mathtt S}_{\beta})$. 
At the end of the day, there is only one symmetric matrix, 
${\rm Im} \mathbf{C}_{{\mathtt S}_{\beta}} = -i \mathbf{C}_{{\mathtt S}_{\beta}}$, 
or its inverse, associated to one ${\mathtt S}_{\beta} \in {\rm WSp}(2n,\mathbb C)$. 
The dimension of the set of symmetric matrices in ${\rm Mat}(2n,\mathbb R)$
is $n(2n+1)$ and is equal to the dimension of ${\rm WSp}(2n,\mathbb C)$, 
since this is the dimension of its Lie algebra $\mathfrak{sp}(2n)$. 
Consequently, apart from the translation $\eta$ and the constant $H_0$,  %
the symbols in (\ref{Eq:WeylSymThermal}-\ref{Eq:WigSymThermal}) 
are uniquely specified by one matrix ${\mathtt S}_{\beta}$. 
This should be compared with the case of the set of 
pure Gaussian states \cite{littlejohn1986}.
One Wigner function of this set is written exactly as (\ref{Eq:WigCharFunc}), 
but replacing  
${\rm Im} \mathbf{C}_{{\mathtt S}_{\beta}} \rightarrow - {\mathsf S \mathsf S}^\top$, 
with $\mathsf S \in {\rm Sp}(2n,\mathbb R)$.
The dimension of the set of Wigner functions (without translations) is $n(n+1)$, 
thus, smaller than the dimension of $\mathfrak{sp}(2n)$. 

Including the translations with $\eta \in \mathbb R^{2n}$ and the constant 
$H_0 \in \mathbb R$, 
the dimension of the set of symbols of a thermal operator is 
$(n+1)(2n+1)$, corresponding to the dimension of ${\rm I}\mathfrak{ m p}(2n)$, 
thus a symbol is completely determined by one Hamiltonian in (\ref{Eq:ClQuadLinHam}).  

For completeness, 
consider now the case of a linear Hamiltonian $\hat H_1 = \hat x \wedge \eta$, 
which generates Weyl operators $\hat T_{\eta t}:= \exp[-i/\hbar \,\hat x \wedge \eta t]$.
From (\ref{Eq:WickThSymb}) and (\ref{Eq:TransRefSymb}), 
the Wigner symbol becomes $E_\beta(x) = \exp[-\beta x \wedge \eta]$. 
Since Wigner and Weyl symbols are related by the 
Fourier transformation (\ref{Eq:SympFour}), 
the Weyl symbol does not exist
[note that such an exponential function $E_\beta(x)$ 
is not absolutely integrable].  
By other side, 
the use of (\ref{Eq:WickThSymb}) for the symbol 
$\tilde T_{\eta t}(\xi)$ from (\ref{Eq:TransRefSymb}) 
generates an ill-defined Dirac delta function.  

%%%%%%%%%%%%%%%%%%%%%%%%%%%%%%%%%%%%%%%%%%%%%%%%%%%%%%%%%%%%%%%%%%%%%%%%%%%%%%%%%%%%%%%%%
\subsection{Covariance of Symbols}                                   \label{Sec:CS} %%%%%
%%%%%%%%%%%%%%%%%%%%%%%%%%%%%%%%%%%%%%%%%%%%%%%%%%%%%%%%%%%%%%%%%%%%%%%%%%%%%%%%%%%%%%%%%
Under the transformation (\ref{Eq:QuadTransf}), 
the symbols of the thermal operator
behaves as (\ref{Eq:SymbsCov}), {\it i.e.}, 
\begin{equation}                                                                         \label{Eq:WickCovRel}
\begin{aligned}
\tilde{E}'_\beta(\xi) &= \frac{ 
      {\rm e}^{  
               -\tfrac{1}{4\hbar} 
                \xi \cdot \mathsf{J}
({\rm Im} \mathbf{C}_{{\mathtt S}'_{\beta}})^{-1}
                          \mathsf{J} \xi -\beta H_0 } }
{ \sqrt{ \det( \mathtt S'_{\beta}  - \mathsf I_{2n} ) }} 
 {\rm e}^{-\frac{i}{\hbar} \xi \wedge (\eta-\zeta)} ,     \\
E'_\beta(x) &=     
\frac{ 2^n \, 
       {\rm e}^{\tfrac{1}{\hbar} (x-{\sf Q}^{-1}\eta + \zeta) \cdot 
                                  {\rm Im} \mathbf{C}_{{\mathtt S}'_{\beta}} 
                                  (x-{\sf Q}^{-1}\eta + \zeta) - \beta H_0} }
     { \sqrt{ \det ( {\mathtt S}'_{\beta} + \mathsf I_{2n} )} },
\end{aligned}
\end{equation}
where 
\begin{equation}                                                                         \label{Eq:CovCompSymp}
\mathtt{S}'_{\beta} = 
{\rm e}^{-i\hbar \beta \mathsf J {\mathsf Q}^\top{\bf H} {\mathsf Q} } = 
{\rm e}^{-i\hbar \beta{\mathsf Q}^{-1} \mathsf J {\bf H} {\mathsf Q} } = 
{\mathsf Q}^{-1} \mathtt{S}_{\beta} {\mathsf Q}. 
 \end{equation}

Noting that 
$ \det ( {\mathtt S}'_{\beta} \pm \mathsf I_{2n}) =  
\det ( {\mathtt S}_{\beta} \pm \mathsf I_{2n})$ and, 
from Eq.(\ref{Eq:CayleyCov}), that 
${\rm Sng}({\rm Im} \mathbf{C}_{{\mathtt S}'_{\beta}}) = 
{\rm Sng}({\rm Im} \mathbf{C}_{{\mathtt S}_{\beta}})$, 
the indexes (\ref{Eq:WickCZ}) are invariant, {\it i.e.}, 
$\nu^{\pm}_{{\mathtt S}'_{\beta}} = \nu^{\pm}_{{\mathtt S}_{\beta}}$. 
%
% \begin{equation}                                                                         \label{Eq:WickInvCz}
% \nu^{\pm}_{{\mathtt S}'_{\beta}} = \nu^{\pm}_{{\mathtt S}_{\beta}}.  
% \end{equation}
%%%%%%%%%%%%%%%%%%%%%%%%%%%%%%%%%%%%%%%%%%%%%%%%%%%%%%%%%%%%%%%%%%%%%%%%%%%%%%%%%%%%%%%%%
\subsection{Partition Function and Indexes}                              %\label{Sec:pfi}
%%%%%%%%%%%%%%%%%%%%%%%%%%%%%%%%%%%%%%%%%%%%%%%%%%%%%%%%%%%%%%%%%%%%%%%%%%%%%%%%%%%%%%%%%
The \hy{PF} in (\ref{Eq:Partfunc}) for the thermal state 
in question becomes 
\begin{equation}                                                                         \label{Eq:PartQuad}
\mathcal Z_{\beta} = 
\frac{{\rm e}^{-\beta H_0}}
     { \sqrt{\det\left({\mathtt S}_{\beta} - \mathsf I_{2n}\right)} },
\end{equation}
which is trivially invariant under the transformation (\ref{Eq:CovCompSymp}), 
since the trace of an operator is invariant under an unitary similarity. 

Due to the equality of the indexes in (\ref{Eq:WickCZ2}), 
both ways of calculation in (\ref{Eq:Partfunc}) are completely equivalent, 
expressing the fact that the trace of an operator is basis-independent. 
Remarkably, $\lim_{\beta \to 0} \mathtt S_{\beta} = \mathsf I_{2n}$ 
is the high temperature limit and, as physically expected, 
the \hy{PF} diverges for any Hamiltonian (\ref{Eq:QQuadLinham}). 
It should be clear that this divergence is not 
a consequence of the chosen representation, 
as the ones which may happen for the symbols in 
(\ref{Eq:WeylSymThermal}) and (\ref{Eq:WigSymThermal}). 
However, note that in the same limit $\tilde{E}_\beta(\xi)$ is not defined, 
while ${E}_\beta(x)$ is a decreasing function due to 
$\mathbf{C}_{{\mathtt S}_{\beta}} \to 0_{2n}$, see Eq.(\ref{Eq:Cayley}).     
Other divergences of (\ref{Eq:PartQuad}) can occur when 
${{\mathtt S}_{\beta}}$ has at least an eigenvalue equal to $+1$,
this reflects structural properties of the system Hamiltonian,
and will be investigated in the examples given in next section.  
Naturally there is no dependence on $\eta$ in Eq.(\ref{Eq:PartQuad}), 
since this term, the linear part of the Hamiltonian (\ref{Eq:QQuadLinham}), 
is only a displacement of the fixed point ``$\hat x =0$'' 
of the quadratic part.

Since $\hat H$ is Hermitian, 
the \hy{PF} is the trace of a positive operator, 
thus $\mathcal Z_\beta > 0$. 
This is an extra requirement to the indexes in (\ref{Eq:WickCZ})
besides Eq.(\ref{Eq:WickCZ2}). 
Practically, in calculations, this can be faced as  
redefining (\ref{Eq:PartQuad}) to $|\mathcal Z_{\beta}|$, see Eq.(\ref{Eq:Prop}).
The indexes in Eq.(\ref{Eq:WickCZ}) and the symbols in 
(\ref{Eq:WeylSymThermal}-\ref{Eq:WigSymThermal}) sustain a double cover 
representation between the group ${\rm WSp}(2n,\mathbb C)$ and the set 
of thermal operators. 
This relation is inherited from the one between 
${\rm Sp}(2n,\mathbb R)$ and ${\rm Mp}(2n)$.

The imposition of the index by the positivity 
of the \hy{PF} has some physical and mathematical consequences. 
As another feature of the \hy{CZ}-index, 
the composition of two \hy{MO}s \cite{littlejohn1986,gossonbook2006} 
is such that 
$\hat M_{\mathsf S_{1}} \hat M_{\mathsf S_{2}} = 
\pm \hat M_{\mathsf S_{1}\mathsf S_{2}}$, but only one sign in ``$\pm$" 
is correct if $\hat M_{\mathsf S_{1}}$, $\hat M_{\mathsf S_{2}}$, 
and their respective \hy{CZ}-index are determined. 
The correct sign of the product is itself given by a \hy{CZ}-index, which is 
a function of the \hy{CZ}-indexes of each \hy{MO} and of all the involved 
symplectic matrices $\mathsf S_{1}, \mathsf S_{2}, \mathsf S_{1}\mathsf S_{2}$ \cite{gossonbook2006}. 
The choice made for the sign of the \hy{PF} inhibits the interpretation of 
the set of thermal operators as a group, since the composition of such two operators  
is not guarantied to be correct without a correct index treatment. 
%
%%%%%%%%%%%%%%%%%%%%%%%%%%%%%%%%%%%%%%%%%%%%%%%%%%%%%%%%%%%%%%%%%%%%%%%%%%%%%%%%%%%%%%%%%
\subsection{Thermodynamical Properties}                                    \label{Sec:TP}
%%%%%%%%%%%%%%%%%%%%%%%%%%%%%%%%%%%%%%%%%%%%%%%%%%%%%%%%%%%%%%%%%%%%%%%%%%%%%%%%%%%%%%%%%
The \hy{PF} is the first step towards the derivation of  
thermodynamical quantities \cite{statphys}. 
To this end, consider the Jacobi formula \cite{horn2013} 
for the derivative of the determinant of an invertible matrix: 
\begin{equation}                                                                         \label{Eq:Jacobi}
d (\det {\bf A}) = 
(\det{\bf A}) \, {\rm Tr}\!\left[ {\bf A}^{-1} d{\bf A}\right]. 
\end{equation}
% It will be also considered that $\nu_{{\mathtt S}_{\beta}} = 0$ 
% in (\ref{Eq:PartQuad}),  
% as well as the symplectic condition in Eq.(\ref{Eq:CompSympInv}). 
% $\mathtt S_{\beta}^{\ast} = \mathtt S_{\beta}^{-1} = 
% \mathsf J \mathtt S_{\beta}^\top \mathsf J^\top$.   
%
This will show that all the thermodynamical functions will depend 
only on the eigenvalues of the Hamiltonian matrix $\mathsf J \mathbf H$ 
or on the respective eigenvalues of $\mathtt S_{\beta}$, 
since $\mathcal Z_{\beta}$ is a function only
of the eigenvalues of $\mathtt S_{\beta}$ through the determinant in (\ref{Eq:PartQuad}).
This determinant also guaranties that $\mathcal Z_{\beta}$
is a product of partitions functions of each eigenvalue of $\mathtt S_{\beta}$, 
thus extensive.  
In the following, the case \hy{(P)} is excluded, 
since $(\mathtt S_{\beta} - \mathsf I_{2n})$ and $\mathsf J \mathbf H$ 
are singular matrices, see Sec.\ref{Sec:PH} for this case.

The Helmholtz free energy of the system is 
\begin{equation}                                                                         \label{Eq:HelmFrEn}
F := - \frac{1}{\beta} \ln \mathcal Z_\beta =    
H_0 + \frac{1}{2\beta} \ln \left|\det(\mathtt S_{\beta} - \mathsf I_{2n})\right|,  
\end{equation}
while the internal energy becomes  
\begin{equation}                                                                         \label{Eq:IntEner}
U :=   
- \frac{\partial}{\partial \beta} \ln {\mathcal Z_\beta}  
%
% = & \, H_0 + \tfrac{\hbar}{2} \, {\rm Tr}\!\left[ {\mathsf J \mathbf H} \, 
% {\rm Im}\,{(\mathtt S_{\beta} - \mathsf I_{2n})^{-1}}\right] \\  
%
= H_0 + \tfrac{\hbar}{4} \, {\rm Tr}\!\left[ {\mathbf H} \, 
{\rm Im} \, {\mathbf C}_{\mathtt S_{\beta}}^{-1}\right].   
\end{equation}
The entropy is readily determined through the use of above formulas: 
$ S =  k_{\rm B}\beta (U-F)$.
Finally,  
the heat capacity is  
\begin{equation}                                                                         \label{Eq:HeatCap}
\begin{aligned}
\!\!C  \!:= k_{\rm B} \beta^2 \!\frac{\partial^2 }{\partial \beta^2}\! 
\ln {\mathcal Z_{\beta}}                 
=  -\tfrac{1}{2}k_{\rm B} \hbar^2\beta^2  
{\rm Tr}\!\left[ \frac{(\mathsf J \mathbf H)^2 
\,\mathtt S_{\beta}}{ 
(\mathtt S_{\beta} - \mathsf I_{2n})^{2}}\right];  
\end{aligned}
\end{equation}
the complex condition (\ref{Eq:CompSympInv}) can be used to certify that 
$C^\ast = C$, thus real. 
Needless to say, the mentioned thermodynamical quantities, 
and any other derived from them, are invariant 
under (\ref{Eq:CovCompSymp}). 
%%%%%%%%%%%%%%%%%%%%%%%%%%%%%%%%%%%%%%%%%%%%%%%%%%%%%%%%%%%%%%%%%%%%%%%%%%%%%%%%%%%%%%%%%
\subsection{Classical Limit}\label{Sec:cl}
%%%%%%%%%%%%%%%%%%%%%%%%%%%%%%%%%%%%%%%%%%%%%%%%%%%%%%%%%%%%%%%%%%%%%%%%%%%%%%%%%%%%%%%%%
It is convenient to write the Hessian in (\ref{Eq:ClQuadLinHam}) as 
${\bf H} = \varpi {\bf H}_\#$, 
where ${\bf H}_\#$ is a dimensionless matrix and $\varpi$ 
a characteristic frequency of the system%
\footnote{%%%%%%%%%%%%%%%%%%%%%%%%%%%%%%%%%%%%%%%%%%%%%%%%%%%%%% 
Assuming that every quantity is measured in the SI,
consider the symplectic matrix 
$\mathsf U = (\sqrt{{\rm Kg}\,{\rm s}^{-1}} \, \mathsf I_{2n}) 
 \oplus (\sqrt{{\rm s}\,{\rm Kg}^{-1}}\,  \mathsf I_{2n} )$,
which corresponds to a change of units.
The transformed vector $x' = \mathsf U x$ is composed by
coordinates and momenta both measured in 
$\sqrt{{\rm Kg \, m^2 s^{-1}}}$, 
{\it i.e.}, both with the same unit of $\sqrt{\hbar}$. 
In this system of units, all the elements of the Hessian $\bf H$ 
in Eq.(\ref{Eq:ClQuadLinHam}) are measured in $\rm s^{-1}$ 
to keep the Hamiltonian in Joules. 
Note that this is necessary to correct write the units of 
the thermodynamical quantities, 
see for instance Eq.(\ref{Eq:IntEner}).}. % 
%%%%%%%%%%%%%%%%%%%%%%%%%%%%%%%%%%%%%%%%%%%%%%%%%%%%%%%%%%%%%%%%%
Thus, it is possible to write 
$H' = \tfrac{\varpi}{2} x'\cdot {\bf H}_\# x'$ 
and by the covariance properties, 
nothing changes except the system of units. 
As matter of simplicity, the symbol $\#$ will be forgot 
and the classical (high temperature) limit becomes simply expressed as 
$\bar \beta := \hbar \varpi \beta \ll 1$. 

In this limit, expanding $\mathtt S_{\beta}$ in (\ref{Eq:CompSymp}), 
there is no even-order corrections to the Cayley matrix (\ref{Eq:Cayley}): 
% \[
% \mathtt S_{\beta} = 
% \mathsf I_{2n} - i \bar \beta \mathsf J \mathbf H
% -\frac{1}{2} \bar \beta^2 (\mathsf J \mathbf H)^2 + 
% \frac{i}{6} \bar \beta^3 (\mathsf J \mathbf H)^3 + \mathcal O^4(\bar \beta)
% \] 
% is justified. Keeping only second order terms and 
\[
\mathbf{C}_{{\mathtt S}_{\beta}} = 
-\frac{i}{2} \bar \beta \mathbf H
+\frac{i}{24} \bar \beta^3 \mathsf J(\mathsf J \mathbf H)^3 
+ \mathcal O(\bar \beta^5),  
\]
since it is an anti-Hermitian matrix, see Eq.(\ref{Eq:CompCayley}).
However, with the help of (\ref{Eq:Jacobi}) 
% the denominator in (\ref{Eq:WeylSymThermal}) becomes
% \[
% \begin{aligned}
% \left[ \det(\mathtt S_{\beta}-\mathsf I_{2n})\right]^{-\tfrac12} &\approx 
% % 
% \frac{i^n \left[ 1 + \tfrac{1}{48}\bar \beta^2 {\rm Tr}(\mathsf J \mathbf H)^2 + \mathcal O^4(\bar \beta) \right] }
%      { \bar \beta^{n} \sqrt{ \det {\bf H} }}, 
% \end{aligned}
% \]
% the inverse of the denominator in (\ref{Eq:WigSymThermal}) becomes  
\[
\left[ \det(\mathtt S_{\beta} + \mathsf I_{2n})\right]^{-\tfrac12} = 
\frac{1}{2^n} 
\left[ 1 - \tfrac{1}{16}\bar \beta^2 {\rm Tr}(\mathsf J \mathbf H)^2 \right] 
+ \mathcal O(\bar \beta^4). 
\]
% Consequently, the Wyel symbol (\ref{Eq:WeylSymThermal}) becomes 
% \[
% \tilde{E}_\beta(\xi) \approx  
% \frac{ 
%       {\exp}\left[{  
%                -\tfrac{1}{2\hbar\bar\beta} 
%               \xi \cdot \mathsf{J} {\bf H}^{-1} \mathsf{J}\xi 
%                -\beta H_0 } -\frac{i}{\hbar} \xi \wedge \eta \right] }
%      {\bar \beta^n \det{\bf H}  
%       \sqrt{\det [\mathsf I_{2n} + 
%         \tfrac{1}{4}\bar \beta^2 (\mathsf J \mathbf H)^2] }} ,  
% \]
Consequently, 
the Wigner symbol (\ref{Eq:WigSymThermal}) approximates to  
\begin{equation}                                                                         \label{Eq:QuadBoltz}
{E}_\beta(x) \approx  
\left[ 1 - \tfrac{\bar \beta^2}{16} {\rm Tr}(\mathsf J \mathbf H)^2 \right]   
\exp\left[{-\beta H_{\rm cl}}\right], 
\end{equation}
which is equal to the Boltzmann factor   
of the classical Hamiltonian (\ref{Eq:ClQuadLinHam}) 
if the second order corrections are discarded.   

In the same limit, 
putting the expansion of $\mathtt S_{\beta}$ in 
(\ref{Eq:IntEner}) and in (\ref{Eq:HeatCap}), 
the internal energy and the heat capacity of the system are given, 
respectively, by  
\begin{equation}                                                                         \label{Eq:EquipTeo}
\begin{aligned}
U &= H_0 + n \beta^{-1} 
        - \frac{ \hbar \varpi }{4!} \, \bar \beta \,    
          {\rm Tr}(\mathsf J \mathbf H)^2 + \mathcal O(\bar\beta^3),  \\ 
C_{v} &= n k_{\rm B} + \frac{k_{\rm B}}{4!}\,\bar\beta^2 \,{\rm Tr}(\mathsf J \mathbf H)^2 
       + \mathcal O(\bar\beta^4). 
 \end{aligned}
\end{equation}
Both above formulas express the equipartition theorem \cite{statphys} 
for a generic \hy{QH} if all higher order corrections in $\bar\beta$ 
are neglected. 

%%%%%%%%%%%%%%%%%%%%%%%%%%%%%%%%%%%%%%%%%%%%%%%%%%%%%%%%%%%%%%%%%%%%%%%%%%%%%%%%%%%%%%%%%
%%%%%%%%%%%%%%%%%%%%%%%%%%%%%%%%%%%%%%%%%%%%%%%%%%%%%%%%%%%%%%%%%%%%%%%%%%%%%%%%%%%%%%%%%
\section{General Hamiltonians}                                             \label{Sec:gh}
%%%%%%%%%%%%%%%%%%%%%%%%%%%%%%%%%%%%%%%%%%%%%%%%%%%%%%%%%%%%%%%%%%%%%%%%%%%%%%%%%%%%%%%%%
%%%%%%%%%%%%%%%%%%%%%%%%%%%%%%%%%%%%%%%%%%%%%%%%%%%%%%%%%%%%%%%%%%%%%%%%%%%%%%%%%%%%%%%%%
The symbols of a thermal operator for a generic Hamiltonian can be obtained using 
the Wick mapping (\ref{Eq:WickThSymb}) of the symbol of the corresponding 
unitary operator. 
However, there are few cases in which the symbol of an unitary operator can be 
analytically settled. Fortunately, semiclassical methods can help in determine
properties of such generic thermal operators.  

The only situations where the (normalized) 
Wigner symbol\footnote{Note the absence of the factor $2^n$ 
in this definition when compared to the one in Eq.(\ref{Eq:WWsymbs}).} 
$\bar H(x) := {\rm Tr}(\hat H \hat R_x)$ 
of a quantum Hamiltonian $\hat H$ is equal to the classical Hamiltonian $H_{\rm cl}$ 
are the quadratic case in (\ref{Eq:QQuadLinham}) \cite{littlejohn1986}, 
the separable case $\hat H = T(\hat p) + V(\hat q)$ \cite{ozorio1998}, 
and either a combination of both.
For a generic quantum system, the (normalized) Wigner symbol of 
the Hamiltonian is a smooth function which 
tends to the classical value in the limit $\hbar \rightarrow 0$ \cite{ozorio1998}. 
The (normalized) Wigner symbol of powers $\hat H^k$ of the Hamiltonian, 
$\bar H^k(x) := {\rm Tr}(\hat H^k \hat R_x)$,
is obtained by the Groenewold formula \cite{groenewold},     
\begin{equation}                                                                         \label{Eq:GroeHam}
\bar H^k(x) = 
\left.
      {\rm e}^{ ^
               - \tfrac{i \hbar}{2} 
                 \tfrac{\partial}{\partial x_1} \wedge 
                 \tfrac{\partial}{\partial x_2} } \, 
                 \bar H(x_1)\bar H^{k-1}(x_2)                      
\right|_{\overset{\scriptstyle x_1 = x}{\scriptstyle x_2 = x}},
\end{equation}
which is no longer factorizable into individual symbols. 
However, since the Weyl symbol of a Hermitian operator is real,  
the recursive expansion of the above exponential shows that
\begin{equation}                                                                         \label{Eq:PotHam}
\bar H^k(x) = [\bar H(x)]^k + \mathcal O(\hbar^2),  
\end{equation}
{\it i.e.}, the symbols of powers of the Hamiltonian differs 
from the power of the symbol only in second order in $\hbar$. 
For $k = 2$, it will be useful to go one 
order further in the expansion of Eq.(\ref{Eq:GroeHam}) to find 
\begin{equation}                                                                         \label{Eq:2PotHam}
\bar H^2(x) = [\bar H(x)]^2 - \frac{\hbar^2}{8} 
{\rm Tr}\!\left(\mathsf J \,\partial^2_{xx} \bar H \right)^2 
+ \mathcal O(\hbar^4).                 
\end{equation}
Note that all odd powers 
in the expansion of (\ref{Eq:GroeHam}) are null,
since the Wigner symbol of any Hermitian operator is real, 
see Eq.(\ref{Eq:HermSymmet}). 

In the crude semiclassical limit, $\hbar \rightarrow 0$, 
square and higher powers of $\hbar$ can be discarded in (\ref{Eq:PotHam}). 
Furthermore, 
the Taylor expansion of the thermal operator ${\rm e}^{-\beta \hat H}$ and 
the linearity of the trace are enough to write 
\begin{equation}                                                                         \label{Eq:ScWigSymb}
\!\!\bar E_\beta(x)  := \!{\rm Tr}({\rm e}^{-\beta \hat H} \!\hat R_x) 
 = \!\sum_{k = 0}^\infty \tfrac{(-\beta)^k}{k!} \bar H^k(x)    
\approx {\rm e}^{-\beta \bar H(x)},
\end{equation}
due to Eq.(\ref{Eq:PotHam}). 
In this limit, the symbol $\bar H(x)$ gets arbitrarily closer to the classical 
Hamiltonian of the system $H_{\rm cl}$, 
which shows that $\bar E_\beta(x)$ 
approaches to the classical Boltzmann factor. 

The Semiclassical approximation in \cite{ozorio1998}, 
which is the core of the one realized here, 
involves the expansion of the unitary operator (\ref{Eq:UnitOp}) 
in powers of $t/\hbar$, 
which restricts the convergence to small time intervals. 
Consequently, 
the composition of Wigner symbols of the unitary operator for small steps  
is necessary to construct an approximation valid for any time. 
Since the expansion for the thermal operator itself in (\ref{Eq:ScWigSymb}) 
does not involves $\hbar$,  
the approximation is valid for any value of $\beta$.

A subtle distinction occurs when including the term in $\hbar^2$ 
from Eq.(\ref{Eq:2PotHam}) into the expansion (\ref{Eq:ScWigSymb}). 
Recalling the unity change in Sec.\ref{Sec:cl}, 
which here is accomplished by $\bar H(x) \to \varpi \bar H(x)$, 
the mentioned term will be the unique contribution of order 
$\bar \beta^2 := \varpi^2\beta^2\hbar^2$ to the sum in (\ref{Eq:ScWigSymb}), 
{\it i.e.}, 
\[
\begin{aligned}
\bar E_\beta(x) &= 
{\rm e}^{-\beta \bar H(x)} - \tfrac{\bar\beta^2}{16} 
{\rm Tr}\!\left(\mathsf J \,\partial^2_{xx} \bar H \right)^2   
+ \mathcal O(\bar\beta^3/\hbar),
\end{aligned}
\]
and can be rewritten as  
\begin{equation}                                                                         \label{Eq:HtWigSymb}
\bar E_\beta(x) =  
\left[1 - \tfrac{\bar\beta^2}{16} 
{\rm Tr}\!\left(\mathsf J \,\partial^2_{xx} \bar H \right)^2 \right]
{\rm e}^{-\beta \bar H(x)} + \mathcal O(\bar\beta^3/\hbar),
\end{equation}
since this last can not be distinguished from the former in 
order of $\beta^3$ \cite{ozorio1998}. 
In the high temperature limit, higher order powers of $\bar \beta$ are discarded, 
and $\bar E_\beta(x)$ approaches the Boltzmann factor 
of the (normalized) Wigner symbol of the Hamiltonian $\hat H$. 
If $\hat H$ is the \hy{QH} in (\ref{Eq:QQuadLinham}), 
then the limit (\ref{Eq:QuadBoltz}) is recovered, 
since $\bar H(x) = {{\rm Tr}( \hat H \hat R_x) }$ is equal to
$H_{\rm cl}$ in (\ref{Eq:ClQuadLinHam}).

The behavior of the system around a critical (fixed) point of the 
symbol $\bar H(x)$ in the semiclassical or in 
the high temperature limits is circumscribed to one 
of those classical categories 
(\hyperlink{(P)}{\color{Gray} P}, 
 \hyperlink{(H)}{\color{Gray} H}, \hyperlink{(E)}{\color{Gray} E} and 
 \hyperlink{(L)}{\color{Gray} L}), 
thus mimicking the examples which will be presented in Sec.~\ref{Sec:Ex}.     
Supposing the existence of $x_0 \in \mathbb R^{2n}$ such that 
$\partial_x \bar H(x) = 0$, 
the ``Hamiltonian'' 
$\bar H(x)$ can be approximated to
\[
\bar H(x) \approx \bar H(x_0) + \tfrac{1}{2} 
                  (x-x_0) \cdot {\bf H}_0 (x-x_0), \,\,\, 
                  \bar {\bf H}_0 := \partial^2_{xx} \bar H(x_0),
\]
for small enough $\delta x:= |x-x_0|$. 
Inserting this expansion in (\ref{Eq:HtWigSymb}), one obtains  
\[
\bar E_\beta(x) \approx 
\left[1 - \tfrac{\bar\beta^2}{16} 
{\rm Tr}\!\left(\mathsf J \bar {\bf H}_0 \right)^2 \right]
{\rm e}^{-\beta H_{\rm cl}^0} + \mathcal O(\bar\beta^3\delta x^3/\hbar),
\]
which is exactly formula (\ref{Eq:QuadBoltz}), since   
\[
H_{\rm cl}^0 := \tfrac{1}{2} x \cdot \mathbf H_0 x + 
                x \wedge x_0  + \bar H(x_0).
\]
  
In principle a Wick rotation could be applied  
to obtain approximations for the thermal operator from the  
well established formulas of semiclassical approximations for unitary operators.
This route was not adopted since some kinds of semiclassical approximations, 
for instance the one in \cite{littlejohn1986,heller}, 
necessarily approximates time-independent Hamiltonians 
by time-dependent ones. 
Such rotation when applied to a time-dependent Hamiltonian 
generates a temperature-dependent one.  

%%%%%%%%%%%%%%%%%%%%%%%%%%%%%%%%%%%%%%%%%%%%%%%%%%%%%%%%%%%%%%%%%%%%%%%%%%%%%%%%%%%%%%%%%
%%%%%%%%%%%%%%%%%%%%%%%%%%%%%%%%%%%%%%%%%%%%%%%%%%%%%%%%%%%%%%%%%%%%%%%%%%%%%%%%%%%%%%%%%
\section{Examples of Quadratic Hamiltonians}                         \label{Sec:Ex} %%%%%
%%%%%%%%%%%%%%%%%%%%%%%%%%%%%%%%%%%%%%%%%%%%%%%%%%%%%%%%%%%%%%%%%%%%%%%%%%%%%%%%%%%%%%%%%
\counterwithin*{equation}{section}
\renewcommand{\theequation}{E-\arabic{equation}}
\setcounter{equation}{0}
%%%%%%%%%%%%%%%%%%%%%%%%%%%%%%%%%%%%%%%%%%%%%%%%%%%%%%%%%%%%%%%%%%%%%%%%%%%%%%%%%%%%%%%%%
The Hamiltonians for the following examples were in majority retrieved 
from the list of normal-form Hamiltonians in \cite{arnold}, 
which is itself a compilation of the results in \cite{williamson1936}. 
All the systems of that list can be worked out in the lines presented here.
However some of them, specially degenerate systems in higher dimensions, 
may require numerical calculations for the determination of the eigenvalues of the 
corresponding Cayley parametrization. 
By another side, this trouble can be circumvented in specific cases, 
see for instance Sec.\ref{Sec:DHH}. 
The determination of the \hy{PF} and/or 
the thermodynamical quantities in Sec.~\ref{Sec:TP} relies only on 
the eigenvalues of $\mathsf J\bf H$, {\it i.e.}, 
the Cayley matrix is not needed for the obtainment of these quantities.  

The term ``normal-form'' indicates the simpler form which a generic quadratic 
Hamiltonian, like (\ref{Eq:ClQuadHam}), can be brought by a symplectic transformation. 
Indeed, for any $\mathsf S \in {\rm Sp}(2n,\mathbb R)$, 
the symplectic transformation $x' = {\mathsf S} x$ changes (\ref{Eq:ClQuadLinHam}) to
\begin{equation}                                                                         \label{Eq:SympTransfHam}
\begin{aligned}
H'_{\rm cl} & = \tfrac{1}{2} x' \cdot {\bf H}  x' + x' \wedge \zeta + H_0 \\
            & = \tfrac{1}{2} {\mathsf S}x \cdot {\bf H} {\mathsf S} x 
              + {\mathsf S}x \wedge \zeta + H_0 \\
            & = \tfrac{1}{2} x \cdot ({\mathsf S}^\top{\bf H} {\mathsf S}) x 
            + x \wedge {\mathsf S}^{-1}\zeta  + H_0, 
\end{aligned}            
\end{equation}
however, the Hamiltonian matrix of the new Hamiltonian $H'_{\rm cl}$ 
becomes similar to the old one   
\begin{equation}                                                                         \label{Eq:SympTransfHam2}
{\mathsf J}{\bf H}' = {\mathsf J} {\mathsf S}^\top {\bf H} {\mathsf S} = 
{\mathsf S}^{-1} (\mathsf J {\bf H}) {\mathsf S}.  
\end{equation}
Thus, following \cite{williamson1936}, 
it is possible to suitably choose 
${\mathsf S}$, such that ${\mathsf J}{\bf H}'$ 
has one of the normal forms in \cite{arnold}.
These normal forms constitute the building blocks, 
which combined generate all \hy{QH}s. 
Note that (\ref{Eq:SympTransfHam}) is accomplished in the quantum case 
by the covariance relations in Sec.\ref{Sec:cos}, 
and the consequences of (\ref{Eq:SympTransfHam}) and (\ref{Eq:SympTransfHam2}) 
are readily translated for thermal states 
as in (\ref{Eq:WickCovRel}) and (\ref{Eq:CovCompSymp}).  

%%%%%%%%%%%%%%%%%%%%%%%%%%%%%%%%%%%%%%%%%%%%%%%%%%%%%%%%%%%%%%%%%%%%%%%%%%%%%%%%%%%%%%%%%
\subsection{Parabolic Hamiltonian}                                   \label{Sec:PH} %%%%%
%%%%%%%%%%%%%%%%%%%%%%%%%%%%%%%%%%%%%%%%%%%%%%%%%%%%%%%%%%%%%%%%%%%%%%%%%%%%%%%%%%%%%%%%%
The most known example of a Hamiltonian in category \hy{(P)} is the free particle.  
One generic Hamiltonian of a $n$ degrees-of-freedom system in \hy{(P)}
is the one having only kinetic energy. 
The Hessian (\ref{Eq:ClQuadHam}) of such a Hamiltonian is      
\begin{equation}                                                                         \label{ExP:Hess}
{\bf H} =  {\bf 0}_n \oplus {\bf M},   
\,\,\, 
{\bf M} \in {\rm Mat}(n,\mathbb R).
\end{equation}
The Hamiltonian matrix and the symplectic matrix for this 
Hamiltonian becomes 
\[
{\mathsf J \bf H} = \left(\begin{array}{cc}
                           {\bf 0}_n & {\bf M} \\
                           {\bf 0}_n & {\bf 0}_n
                          \end{array}      \right), 
\,\,\,
\mathsf S_{t} = \mathsf I_{2n} + {\mathsf J \bf H} t = 
                \left(\begin{array}{cc}
                       \mathsf I_{n} & \mathbf M t \\
                       \mathbf 0_{n} & \mathsf I_{n}
                      \end{array}      \right).
\]
Note that eigenvalues of ${\mathsf J \bf H}$ are all null 
and the ones of $\mathsf S_{t}$ are all one, which justify 
the categorization of this system as \hy{(P)}, thus
\begin{equation}                                                                         \label{ExP:Det}
\det(\mathsf S_{t} - \mathsf I_{2})  = 0, \,\,\,  
\det(\mathsf S_{t} + \mathsf I_{2})  = 2^{2n} \,\,\,\,\, (\forall t >0). 
\end{equation}
Consequently, there is no divergence for 
the Wigner representation (\ref{Eq:Hsmet}) of the metaplectic operator, 
while there is absolutely no Weyl representation (\ref{Eq:Wsmet}).  
The Cayley parametrization (\ref{Eq:Cayley}) becomes
\[%\begin{equation}                                                                         \label{ExP:Cayley}
{\bf C}_{\mathsf S_{t}} = {\bf 0}_n \oplus \tfrac{1}{2}{\bf M} t
\]%\end{equation}
and the Eq.(\ref{Eq:CZlimit}) sets  
$\nu^{+}_{\mathsf{S}_{t}}(t) = 0, \, \forall t > 0$. 
Collecting all these results, the Wigner symbol for 
the metaplectic operator associated to the free particle Hamiltonian is 
\begin{equation}                                                                         \label{ExP:WigSymb}
M_{\mathsf S_{t}}(x) = \exp\left[- \frac{i t}{2\hbar} p\cdot {\bf M} p \right].  
\end{equation}

Considering the thermal operator for the same Hamiltonian, 
the complex symplectic matrix in (\ref{Eq:CompSymp}) becomes
\[
\mathtt S_{\beta} = \mathsf I_{2n} -i \hbar \beta {\mathsf J \bf H} .   
\]
Since the eigenvalues are preserved by the Wick mapping, see Eq.(\ref{Tab:WickClass}), 
$\det(\mathtt S_{\beta} - \mathsf I_{2})  = 2^{2n}$, $\forall t >0$,
as in (\ref{ExP:Det}).  
The Cayley parametrization (\ref{Eq:Cayley}) for $\mathtt S_{\beta}$ is 
\[%\begin{equation}                                                                         \label{ExP:WickCayley}
{\bf C}_{\mathtt S_{\beta}} = {\bf 0}_n \oplus 
(- \tfrac{i}{2}{\bf M} \hbar\beta),  
\]%\end{equation}
and, accordingly to (\ref{Eq:WCZlimit}), $\nu^{+}_{{\mathtt S}_{\beta}} = 0$. 

The Wigner symbol in (\ref{Eq:WigSymThermal}) becomes 
\[
E_\beta(x) = \exp\left[- \frac{\beta}{2} p\cdot {\bf M} p \right], 
\]
which is simply the Wick mapping (\ref{Eq:WickRot}) applied to (\ref{ExP:WigSymb}).   

Since $\det(\mathtt S_{\beta} - \mathsf I_{2})  = 0$, $\forall t >0$, 
there is no Weyl representation for the thermal operator, as before.  
However, the matrix ${\rm Im}{\bf C}_{\mathtt S_{\beta}}$ is negative-semidefinite 
for a positive-definite $\bf M$. 
Pushing the luck, this is enough to guarantee the convergence 
of the Fourier integral (\ref{Eq:SympFour}) of the Wigner symbol 
in the momentum space, while in coordinate space, a delta distribution is used: 
\[
\begin{aligned}
\tilde E_\beta(\xi) & = 
\int_{{\mathbb R}^n} \!\! 
{\rm d}^n\!q \, \frac{ {\rm e}^{-\frac{i}{\hbar} q\cdot \xi_p}}{(2\pi\hbar)^n} 
\int_{{\mathbb R}^n} \!\! 
{\rm d}^n\!p \,\,  {\rm e}^{- \frac{\beta}{2} p\cdot {\bf M} p 
                                        + \frac{i}{\hbar} p\cdot \xi_q}    \\
& = \frac{(2\pi)^n}{\beta^n \sqrt{\det \mathbf M}} \, \delta^n\!(\xi_p) \, 
    {\exp}\!\left[ -\frac{\xi_q \cdot {\bf M}^{-1} \xi_q    }{2\hbar^2 \beta} \right] . 
\end{aligned}
\]

The \hy{PF} in (\ref{Eq:PartQuad}) is divergent, 
since $\det(\mathsf S_{t} - \mathsf I_{2})  = 0$. 
Note that, from above Weyl symbol, $\tilde E_\beta(0)$ 
is a delta-function.  
However, it is customary in statistical physics to force a 
convergence by truncation of integrals like the one in (\ref{Eq:Partfunc}).
This is accomplished for the presented example considering the integral in 
configuration space as the volume $\mathcal V$ 
occupied by the system:    
\begin{equation}                                                                         \label{ExP:partfunc}
\begin{aligned}
\mathcal Z_\beta = \int_{\mathbb R^n}\!\!{\rm d}^{n}\!q 
                   \int_{\mathbb R^n}\!\!{\rm d}^{n}\!p \frac{E_\beta(x)}{(2\pi\hbar )^n} 
 = \frac{\mathcal V}{(2 \pi\hbar^2\beta)^\frac{n}{2} \sqrt{\det \mathbf M}}. 
\end{aligned}
\end{equation}

Consider now a rotation in phase space $x' = \mathsf J x$. 
The new Hamiltonian $\hat H' = \tfrac{1}{2} q \cdot \mathbf M q$
has Hessian  
\[
{\bf H}' = {\mathsf J}^\top{\bf H}{\mathsf J}  =   {\bf M} \oplus {\bf 0}_n.    
\]
From the covariance relation (\ref{Eq:WickCovRel}), 
the Wigner symbol in (\ref{Eq:WigSymThermal}) becomes 
\[
E'_\beta(x) = \exp\left[- \frac{\beta}{2\hbar} q\cdot {\bf M} q \right], 
\]
with $\nu^{+}_{{\mathtt S}_{\beta}} = \nu^{+}_{{\mathtt S}'_{\beta}} = 0$, 
see Sec.\ref{Sec:CS}. 
The Fourier transformation of above symbol gives
\[
\tilde E'_\beta(\xi)  = 
\frac{(2\pi)^n}{\beta^n \sqrt{\det \mathbf M}} \, \delta^n\!(\xi_q) \, 
{\exp}\!\left[ -\frac{\xi_p \cdot {\bf M}^{-1} \xi_p}{2\hbar^2 \beta} \right] . 
\]
The \hy{PF} is again obtained by integrating the Wigner symbol 
$E'_\beta(x)$, however the convergence now is performed truncating 
the momenta-integral, and becomes
\[
\begin{aligned}
\mathcal Z_\beta = \int_{\mathbb R^n}\!\!{\rm d}^{n}\!p 
                   \int_{\mathbb R^n}\!\!{\rm d}^{n}\!q \frac{E'_\beta(x)}{(2\pi\hbar )^n} 
 = \frac{\mathcal V'}{(2 \pi\hbar^2\beta)^\frac{n}{2} \sqrt{\det \mathbf M}}, 
\end{aligned}
\]
where $\mathcal V'$ is the volume of momenta space. 
Due to the truncation performed, $\mathcal Z_\beta$ is not invariant under 
the symplectic transformation $x'= \mathsf J x$, as it should be.

Concluding, the procedure to obtain the symbols and the thermodynamical 
properties of system in category \hy{(P)} departs from the obtainment of 
the symbol $E_\beta(x)$, and then the \hy{PF} through 
a truncation of the integrals, 
which can be independently either on momenta or coordinates, 
or even in a mixture of both. 
Formulas in Sec.\ref{Sec:TP} do not work, 
and the thermodynamical functions should be derived directly for 
\hy{PF}s as in the standard literature \cite{statphys}. 
%

%%%%%%%%%%%%%%%%%%%%%%%%%%%%%%%%%%%%%%%%%%%%%%%%%%%%%%%%%%%%%%%%%%%%%%%%%%%%%%%%%%%%%%%%%
\subsection{Hyperbolic Hamiltonian}                                 \label{Sec:hh}  %%%%%
%%%%%%%%%%%%%%%%%%%%%%%%%%%%%%%%%%%%%%%%%%%%%%%%%%%%%%%%%%%%%%%%%%%%%%%%%%%%%%%%%%%%%%%%%
The one degree of freedom normal-form Hamiltonian in category \hy{(H)} is 
$H_{\rm cl}= \kappa p q$ \cite{arnold}. 
In quantum optics, its symmetric quantized version, 
$\hat H = \tfrac{\kappa}{2} (\hat q \hat p + \hat p \hat q)$, 
is responsible for the phenomenon of squeezing \cite{QuantOptics}.  

The Hessian, the Hamiltonian matrix and the symplectic matrix for 
the Hamiltonian are 
\[
{\bf H} =  \kappa {\bm \sigma}_{\!\tt x}, 
\,\, 
{\mathsf J \bf H} = \kappa {\bm \sigma}_{\!\tt z}, \,\,
%           \begin{pmatrix}
%            \kappa & 0 \\
%            0 & -\kappa
%            \end{pmatrix}, 
\mathsf S_{t} = {\rm Diag}({\rm e}^{\kappa t}, {\rm e}^{-\kappa t}), 
\]
where ${\bm \sigma}_{\!\tt x}$ and ${\bm \sigma}_{\!\tt z}$ 
are the Pauli matrices. 
The complex symplectic matrix (\ref{Eq:CompSymp}) is, thus, 
\[
\mathtt S_{\beta} ={\rm Diag}({\rm e}^{-i\bar \beta }, {\rm e}^{i\bar \beta}), 
\,\,\,
\bar \beta := \hbar \kappa \beta.  
\]
As stated in (\ref{Tab:WickClass}), 
the eigenvalues of $\mathsf J \bf H$ (and of $\mathsf S_{t}$) classifies 
this matrix as \hy{(H)}, and the Wick rotation generates a matrix  
$\mathtt S_{\beta}$ in \hy{(E)}.  

The Cayley parametrization (\ref{Eq:Cayley}) for $\mathtt S_{\beta}$ is 
\begin{equation}                                                                         \label{ExH:WickCayley}
{\bf C}_{\mathtt S_{\beta}} = - i \, 
{\rm tg}\!\left(\tfrac{\bar \beta}{2}\right) {\bm \sigma_{\tt x}},  
\end{equation}
and
\begin{equation}                                                                         \label{ExH:Det}
\begin{aligned}
\det(\mathtt S_{\beta} - \mathsf I_{2})  = 
4\sin^2\!\left(\tfrac{\bar \beta}{2}\right), \\  
\det(\mathtt S_{\beta} + \mathsf I_{2})  = 
4\cos^2\!\left(\tfrac{\bar \beta}{2}\right). 
\end{aligned}
\end{equation}

The Weyl symbol in (\ref{Eq:WeylSymThermal}), 
using Eqs.(\ref{Eq:WickCZ}), (\ref{ExH:WickCayley}) and (\ref{ExH:Det}), 
becomes
\begin{equation}                                                                         \label{ExH:WeylWickSymb}
\tilde E_\beta(\xi) = 
\frac{i^{\nu^-_{\mathtt{S}_\beta}}}{2} 
\left| {\rm csc} \! \left( \tfrac{\bar \beta}{2} \right) \right|    
{\exp}\left[{ - \frac{1}{2\hbar} \, 
          {\rm ctg}\!\left( \tfrac{\bar \beta}{2} \right) 
          \xi_q \, \xi_p } \right],
\end{equation}
and is not defined when $\bar \beta = \hbar \kappa \beta = m\pi$, 
for $m \in \mathbb N$.
The Wigner symbol (\ref{Eq:WigSymThermal}) is expressed as
\begin{equation}                                                                         \label{ExH:WigWickSymb}
% E_\beta(x) = 
% \frac{ i^{\nu^+_{\mathtt{S}_\beta} }
%        \exp\left[ - \frac{2}{\hbar} \, 
%                     {\rm tg}\!\left( \tfrac{\hbar \kappa \beta}{2} \right) q \, p \right] } 
%      { \left| \cos \! \left( \tfrac{\hbar \kappa \beta}{2} \right) \right| },
E_\beta(x) = 
i^{\nu^+_{\mathtt{S}_\beta}}
\left| {\rm sec} \! \left( \tfrac{\bar \beta}{2} \right) \right| 
{\exp}\left[ - \frac{2}{\hbar} \, 
              {\rm tg}\!\left(\tfrac{\bar \beta}{2} \right) q \, p \right],
\end{equation}
and is not defined when $\bar \beta = \hbar \kappa \beta = (2m+1)\pi$, 
for $m \in \mathbb N$.   
The \hy{PF} ({\ref{Eq:PartQuad}}) obtained through $\tilde E_\beta(0)$ is
\begin{equation}                                                                         \label{ExH:PartFun}
\mathcal Z_{\beta} = \frac{i^{\nu^-_{\mathtt{S}_\beta}}}{2} 
                     \left|{\rm csc}\left(  \tfrac{\bar\beta}{2} \right)\right|. 
\end{equation}

It remains to determine the indexes
in (\ref{ExH:WeylWickSymb}), in (\ref{ExH:WigWickSymb}), and in (\ref{ExH:PartFun}).  
From Eq.(\ref{ExH:WickCayley}),  
${\rm Spec}({\rm Im} \mathbf{C}_{{\mathtt S}_{\beta}}) = 
\{\pm {\rm tg}({\hbar \kappa \beta}/{2})\}$, consequently   
neither (\ref{ExH:WeylWickSymb}) nor (\ref{ExH:WigWickSymb})
are (absolutely) integrable functions, 
which inhibits the Fourier transformation (\ref{Eq:SympFour}) among the symbols. 
However, the \hy{PF} is the trace of a positive operator, 
thus it is positive, which gives $\nu^-_{\mathtt{S}_\beta} = 0$, 
$\forall \beta \ge 0$. 
From (\ref{Eq:WCZlimit}), $\lim_{\beta \to 0}E_\beta(x) = 1$ and thus, 
taking into account the first divergence of (\ref{ExH:WigWickSymb}),  
$\nu^+_{\mathtt{S}_\beta} = 0$ for $0 \le \hbar \kappa \beta < \pi $, 
and this is the only interval where it is possible to determine 
$\nu^+_{\mathtt{S}_\beta}$, 
due to the absence of a Fourier transformation.   
By the same reasons, the Wigner function (\ref{Eq:WigCharFunc}) 
for the thermal state of the hyperbolic Hamiltonian is not defined,
but the chord function is given also in 
(\ref{Eq:WigCharFunc}) with (\ref{ExH:WickCayley}). 

\begin{figure}[t]
\includegraphics[width=7.5cm, trim=0 15 0 0]{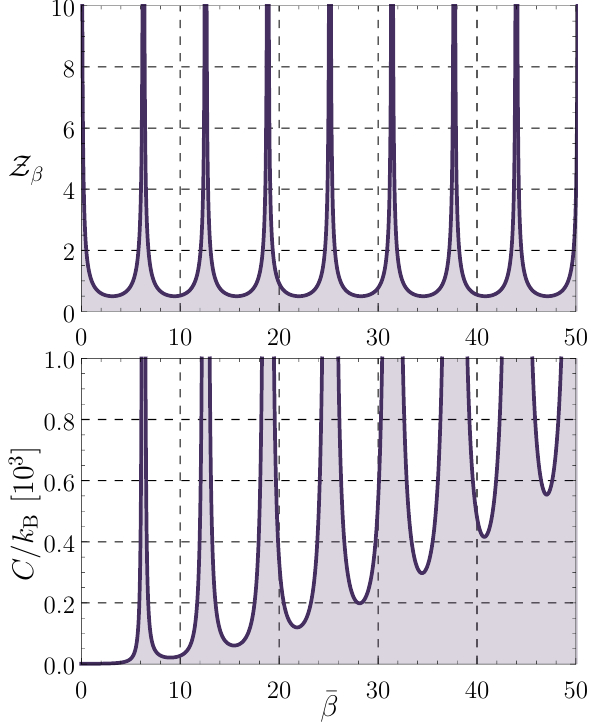}
\caption{ Partition function (top) and the Heat Capacity (bottom) 
of the Thermal State for the one degree of freedom 
Hyperbolic Hamiltonian $H(x) = \kappa pq$ 
as a function of the inverse temperature 
$\bar\beta := \hbar \kappa \beta$.}                                        \label{Fig:H}
\end{figure}

The heat capacity for this Hamiltonian, from (\ref{Eq:HeatCap}), is   
\[
C = \tfrac{1}{4} k_{\rm B} \bar \beta^2 {\rm csc}^2\!\left(\frac{\bar\beta}{2}\right), 
\]
and is plotted in Fig.\ref{Fig:H}, 
where it is also shown the \hy{PF} of the system. 
Both functions diverges for 
$\bar \beta = \bar\beta_{\rm c} = 2 m \pi$, 
and expanding the above formula around these values, 
one obtains a critical exponent $\alpha = 2$. 
However, in the present situation, this divergence is a 
pure mechanical effect and cannot be faced as a kind of phase transition, 
since the internal energy (\ref{Eq:IntEner}) and 
the free-energy (\ref{Eq:HelmFrEn}) are themselves discontinuous 
at the same points. 
Note that $\lim_{\beta \to 0} C = k_{\rm B}$ in agreement with 
(\ref{Eq:EquipTeo}). 

The scattering of a particle through a parabolic barrier is described by 
inverted oscillator $H' = \tfrac{\kappa}{2}(p^2 - q^2)$, 
which is a rotation of the hyperbolic Hamiltonian considered,
{\it i.e.}, $H'_{\rm cl} = H_{\rm cl}({\mathsf R} x)$, where 
\[
\mathsf R = \frac{1}{\sqrt{2}}\begin{pmatrix}
                               1 & -1 \\
                               1 &  1  
                              \end{pmatrix} \in {\rm Sp}(2,\mathbb R).
\]
The symbols of the thermal operator for the inverted oscillator are readily obtained
using the covariance relations (\ref{Eq:WickCovRel}), while the \hy{PF}, 
since it is invariant, is the same as (\ref{ExH:PartFun}).

As an observation, similar to what was done in Eq.(\ref{ExP:partfunc}) 
for the configuration space, 
an attempt to truncate the integration of (\ref{ExH:WigWickSymb}) 
in position and either in momentum,  
constraining the system to a phase space volume $\Omega$, attain  
\begin{equation}                                                                         \label{ExH:parttrunc}
\mathcal Z_{\beta} = 
\frac{1}{\pi} 
\left|{\rm csc}\left(  \tfrac{\bar\beta}{2} \right)\right|\,
{\rm Shi}\! 
\left[ \frac{2\Omega}{\hbar} 
       {\rm tg}\!\left(  \tfrac{\bar\beta}{2} \right) \right], 
\end{equation}
and does not remove the divergences of the \hy{PF}.  

As a last comment, 
it should be also noted that, since all the thermodynamical 
quantities are functions exclusively of the eigenvalues 
of $\mathtt S_\beta$, or the ones of $\mathsf J \bf H$, 
their behavior is a property of the whole class of systems in 
category \hy{(H)}.  

%%%%%%%%%%%%%%%%%%%%%%%%%%%%%%%%%%%%%%%%%%%%%%%%%%%%%%%%%%%%%%%%%%%%%
%%%%%%%%%%%%%%%%%%%%%%%%%%%%%%%%%%%%%%%%%%%%%%%%%%%%%%%%%%%%%%%%%%%%%
%%%%%%%%%%%%%%%%%%%%%%%%%%%%%%%%%%%%%%%%%%%%%%%%%%%%%%%%%%%%%%%%%%%%%
%%%%%%%%%%%%%%%%%%%%%%%%%%%%%%%%%%%%%%%%%%%%%%%%%%%%%%%%%%%%%%%%%%%%%
%%%%%%%%%%%%%%%%%%%%%%%%%%%%%%%%%%%%%%%%%%%%%%%%%%%%%%%%%%%%%%%%%%%%%
%%%%%%%%%%%%%%%%%%%%%%%%%%%%%%%%%%%%%%%%%%%%%%%%%%%%%%%%%%%%%%%%%%%%%

%%%%%%%%%%%%%%%%%%%%%%%%%%%%%%%%%%%%%%%%%%%%%%%%%%%%%%%%%%%%%%%%%%%%%%%%%%%%%%%%%%%%%%%%%
\subsection{Degenerate Hyperbolic Hamiltonian}                      \label{Sec:DHH} %%%%%
%%%%%%%%%%%%%%%%%%%%%%%%%%%%%%%%%%%%%%%%%%%%%%%%%%%%%%%%%%%%%%%%%%%%%%%%%%%%%%%%%%%%%%%%%
From the generality of the \hy{QH}s, 
it is interesting to work on a nontrivial system dynamics. 
The normal-form Hamiltonian of a system with $n$ degrees of freedom 
and $n$-fold degenerate pairs of eigenvalues $(\kappa,-\kappa)$ in category \hy{(H)} 
is \cite{arnold} 
\[%begin{equation}                                                                         \label{ExDH:Ham}
H_{\rm cl} = \kappa \sum_{j=1}^n p_j q_j - \kappa\sum_{j=1}^{n-1} p_j q_{j+1},   
\]%end{equation}
which reduces to the case of previous example for $n=1$.  
The Hessian of the Hamiltonian is 
\begin{equation}                                                                         \label{ExDH:Hess}
{\bf H} =  \left( \begin{array}{cc}
            {\bf 0}_n    &  {\bf K}  \\
            {\bf K}^\top &  {\bf 0}_n
           \end{array}                    \right), 
\,\,\, 
{\bf K} := \kappa (\mathsf I_{n} - {\bf N}),  
\end{equation}
and ${\bf N}\in {\rm Mat}(n,\mathbb R)$ is 
the nilpotent matrix with entries ${\bf N}_{jl} := \delta_{j,l+1}$. 
The Hamiltonian matrix becomes
\[ 
\mathsf J {\bf H} = {\bf K}^\top \oplus (-{\bf K}),  
\]
and generates the complex symplectic matrix, {\it via} (\ref{Eq:CompSymp}),  
\begin{equation}                                                                         \label{ExDH:CompSymp}
\mathtt S_{\beta} = ({\rm e}^{-i\bar\beta} {\rm e}^{i\bar \beta {\bf N}^\top}) 
                     \oplus 
                    ({\rm e}^{ i\bar\beta} {\rm e}^{-i\bar \beta {\bf N}}), 
\,\,\,
\bar\beta := \hbar \kappa \beta, 
\end{equation}
where 
\[ 
{\rm e}^{{\bf N} \alpha} = \sum_{j=0}^{n-1} \frac{\alpha^j}{j!} {\bf N}^j =     
\left( \begin{array}{ccccc}
        1      & \alpha      & \tfrac{\alpha^2}{2} & \cdots & \frac{\alpha^{n-1}}{({n-1})!} \\
        0      & 1      & \alpha         & \ddots & \vdots \\
        0      & 0      & 1         & \ddots & \tfrac{\alpha^2}{2} \\
        \vdots & \ddots & \ddots         & \ddots & \alpha \\
        0      & \cdots & 0              & 0      & 1
       \end{array}                    \right). 
\]
Due to the triangular structure of the blocks of $\mathtt S_{\beta}$, 
it is not difficult to show that its $n$-fold degenerate spectrum  
is $\{\exp(-i\bar\beta),\exp(i\bar\beta)\}$, thus 
\[
\begin{aligned}
\det(\mathtt S_{\beta} - \mathsf I_{2n}) &= 
2^{2n} \sin^{2n}\!\left( \tfrac{\bar \beta}{2}\right),   \\
\det(\mathtt S_{\beta} + \mathsf I_{2n}) &= 
2^{2n} \cos^{2n}\!\left( \tfrac{\bar\beta}{2}\right).   
\end{aligned}
\]
From these two, the symbols $\tilde E_\beta(\xi)$ and $E_\beta(\xi)$  
diverges, respectively, 
as in (\ref{ExH:WeylWickSymb}) and as in (\ref{ExH:WigWickSymb}). 

An explicit expression for the Cayley parametrization (\ref{Eq:Cayley}) 
of $\mathtt S_{\beta}$ involves lots of cumbersome expressions,
however, it is possible to write 
\begin{equation}                                                                          \label{ExDH:WickCayley}
{\bf C}_{\mathtt S_{\beta}} = \begin{pmatrix}
                              {\bf 0}_n & -[F({\bf N})]^\ast\\
                              F({\bf N}^\top)  & {\bf 0}_n 
                              \end{pmatrix}.
%  
% \mathsf J [F({\bf N}) \oplus F({\bf N}^\top)],   
\end{equation}
Taking advantage of the nilpotency of $\mathbf N$, 
the function $F$ can be represented as a finite power series:   
\[
\begin{aligned}
F({\bf N}) & := 
[ {\rm e}^{ i\bar \beta {\bf N}  }  + {\rm e}^{ i\bar \beta} \mathsf I_{n} ]^{-1}
[{\rm e}^{ i\bar \beta {\bf N}  }  - {\rm e}^{ i\bar \beta} \mathsf I_{n} ]        \\
 &= \sum_{m = 0}^{n-1} \frac{{\bf N}^m}{m!} 
                       \frac{\partial^m }{\partial {\bf N}^m} 
                       F({\bf N}),  
\end{aligned}
\]
which can be easily computed in any symbolic computational software.  
Since a general analytic expression for ${\bf C}_{\mathtt S_{\beta}}$ is missing, 
the determination of the eigenvalues of ${\rm Im}{\bf C}_{\mathtt S_{\beta}}$ 
is impossible. 
However, its characteristic polynomial, from (\ref{ExDH:WickCayley}), is
\[
\begin{aligned}
P(\lambda) &= 
\det\left[ \lambda \mathsf I_{2n} - 
           {\rm Im}{\bf C}_{\mathtt S_{\beta}}\right] \\ 
&=
\det\left[ \lambda^2 \mathsf I_n - 
          {\rm Im}F({\bf N}^\top){\rm Im}F({\bf N})  \right]                        
= P(-\lambda),
\end{aligned}
\]
which shows that the Fourier transformation (\ref{Eq:SympFour}) does not converge 
for the symbols (\ref{ExH:WeylWickSymb}) and in (\ref{ExH:WigWickSymb}), 
exactly as in previous example.
Despite of these, the \hy{PF} can be calculated, 
since the Weyl symbol (\ref{ExH:WeylWickSymb}) is well defined, 
which gives, for $\nu^-_{\tt S_{\beta}} = 0$, 
\[
\mathcal Z_{\beta} = 
\frac{1}{2^n} 
\left| {\rm csc}\left( \frac{\hbar \kappa \beta}{2} \right) \right|^n,
\]
which is the \hy{PF} of $n$ non-interacting  
unidimensional systems described by (\ref{ExH:PartFun}).

For illustration, consider the case $n=2$. 
The Hamiltonian is 
$H(x) = \kappa (p_1 q_1 + p_2 q_2 - p_1 q_2 )$  
and the Hessian (\ref{ExDH:Hess}) becomes 
\[
{\bf H} = \left( \begin{array}{cc}
                  {\bf 0}_2      & {\bf K}  \\
                  {\bf K}^\top   & {\bf 0}_2  
                 \end{array}\right), 
\,\,\, 
{\bf K} :=   \left( \begin{array}{cc}
                     \kappa  &  0  \\
                   - \kappa  &  \kappa  
                 \end{array}\right). 
\]
The complex symplectic matrix in (\ref{ExDH:CompSymp}) is  
\[
\mathtt S_{\beta} = 
\left( \begin{array}{cc}
       1 & 0     \\
       i \bar\beta &  1
       \end{array} \right) {\rm e}^{-i \bar \beta }
\oplus
\left( \begin{array}{cc}
       1 & -i \bar\beta    \\
       0 &  1
       \end{array} \right) {\rm e}^{i \bar \beta },  
\,\,\, 
\bar \beta:= \hbar \kappa \beta,  
\]
and the Cayley parametrization (\ref{ExDH:WickCayley}) is written with   
\[
F({\bf N}) = i \left( \begin{array}{cc}
           {\rm tg}(\tfrac{\bar\beta }{2}) &  0  \\
           \frac{\bar\beta }{2} {\rm csc}^2(\tfrac{\bar\beta }{2})  
           & {\rm tg}(\tfrac{\bar\beta }{2})   
           \end{array}\right),  
\]
from where 
the four eigenvalues of ${\rm Im}{\bf C}_{\mathtt S_{\beta}}$ are obtained, 
\[
{\rm Spec}({\rm Im}{\bf C}_{\mathtt S_{\beta}}) = 
\left\{
 \frac{ \pm \bar \beta \pm 
 \sqrt{\bar\beta^2 + 4\sin^2\bar\beta}}{ 2 + 2\cos\bar\beta}\right\},
\]
and constitutes pairs of symmetric values.  
%
%%%%%%%%%%%%%%%%%%%%%%%%%%%%%%%%%%%%%%%%%%%%%%%%%%%%%%%%%%%%%%%%%%%%%%%%%%%%%%%%%%%%%%%%%
\subsection{Elliptic Hamiltonian}                                  \label{Sec:EH} %%%%%
%%%%%%%%%%%%%%%%%%%%%%%%%%%%%%%%%%%%%%%%%%%%%%%%%%%%%%%%%%%%%%%%%%%%%%%%%%%%%%%%%%%%%%%%%
The one degree of freedom harmonic oscillator is the textbook example 
for the Wick mapping and by this reason
deserves attention in the picture presented here.

The Hessian of the Hamiltonian and the Hamiltonian matrix are, respectively,  
${\bf H} = \omega \mathsf I_{2}$ and   
$\mathsf J{\bf H} = \omega \mathsf J$. 
The complex symplectic matrix (\ref{Eq:CompSymp}) becomes
\[
\mathtt{S}_\beta = \begin{pmatrix}
                   \cosh\bar\beta & -i \sinh\bar\beta \\
                 i \sinh\bar\beta & \cosh\bar\beta
                   \end{pmatrix},
\,\,\,
\bar \beta := \hbar \omega \beta;
\]
its Cayley parametrization (\ref{Eq:Cayley}) is
\[
\begin{aligned}
{\bf C}_{\mathtt S_{\beta}} = 
-i \, {\rm tgh}\!\left( \tfrac{\bar\beta}{2}\right) \mathsf I_{2},
\end{aligned}
\]
and 
\[
\begin{aligned}
\det(\mathtt S_{\beta} - \mathsf I_{2n}) & = 
-4 \sinh^{2}\!\left( \tfrac{\bar\beta}{2}\right),   \\
\det(\mathtt S_{\beta} + \mathsf I_{2n}) &= 
4 \cosh^{2}\!\left( \tfrac{\bar\beta}{2}\right) .   
\end{aligned}
\]

From the above determinants, the symbol $\tilde E_\beta(\xi)$ 
is not defined only for $\beta = 0$, while the limit (\ref{Eq:WCZlimit}) 
imposes $\nu^+_{\tt S_{\beta}} = 0$ for the same value of $\beta$. 
Also note that there are no other divergences for both symbols, 
in such a way that the symplectic Fourier transform (\ref{Eq:SympFour})
interchanges both representations at all with the indexes given by (\ref{Eq:WickCZ2}), 
since ${\rm Im}  {\bf C}_{\mathtt S_{\beta}} < 0$, $\forall \beta > 0$.  

The Weyl and Wigner symbols becomes, respectively,
\begin{equation}                                                                         \label{ExE:WickSymbs}
\begin{aligned}
\tilde E_\beta(\xi) &= 
\frac{ \exp\!\left[ -\frac{1}{4} 
                  { \rm ctgh }\!\left(\tfrac{\bar\beta}{2}\right) 
                  \xi^2                                                    \right] }
     { 2 \sinh\!\left( \tfrac{\bar\beta}{2} \right) },                      \\
E_\beta(x) &= 
\frac{ \exp\!\left[ -\frac{1}{\hbar} 
                  { \rm tgh }\!\left(\tfrac{\bar\beta}{2}\right) 
                  x^2                                                    \right] }
     { 2 \cosh\!\left( \tfrac{\bar\beta}{2} \right) }.                      \\
\end{aligned}
\end{equation}
and the \hy{PF} ({\ref{Eq:PartQuad}}) becomes
\begin{equation}                                                                         \label{ExE:PartFun}
\mathcal Z_{\beta} = 
\frac{1}{2} {\rm csch}\! \left( \tfrac{\bar\beta}{2} \right). 
\end{equation}

The widely known ``Williamson theorem'' \cite{Note2} 
is strictly connected with the elliptic case. 
Let $\mathbf M \in {\rm Mat}(2n,\mathbb R)$ 
be any symmetric positive definite matrix: 
$\mathbf M = \mathbf M^\top > 0$. 
The theorem states that this matrix can be diagonalized by 
a symplectic congruence, {\it i.e.}, there exists 
$\mathsf S_\mathbf{M} \in {\rm Sp}(2n, \mathbb R)$ 
such that 
\begin{equation}                                                                         \label{ExE:tw1}      
\mathsf S^\top_\mathbf{M} \mathbf M \mathsf S_\mathbf{M} 
= \Lambda_\mathbf{M} \oplus \Lambda_\mathbf{M}, \,\,\,  
\Lambda_\mathbf{M} := 
{\rm Diag}(\mu_1,...,\mu_n)  
\end{equation}
with $\mu_j > 0$. 
The diagonal matrix 
$\Lambda_\mathbf{M}$ is called {\it symplectic spectrum} of 
$\mathbf M$ and $\mu_i$ the symplectic eigenvalues. 
These can be found to be the (euclidean) 
eigenvalues of $\mathsf J \mathbf M$, {\it i.e.}, 
\begin{equation}\label{ExE:tw2}
{\rm Spec_{\mathbb C}}(\mathsf J \mathbf M) = 
{\rm Diag}(i\mu_1,...,i\mu_n, -i\mu_1,...,-i\mu_n).
\end{equation}

Thus, any positive \hy{QH}, 
{\it i.e.}, one such that 
$H_{\rm cl} = \tfrac{1}{2} x \cdot {\bf M} x$ with $\mathbf M > 0$,
has the collection of harmonic oscillators,
with frequencies given by the symplectic spectrum $\Lambda_{\bf M}$, 
as a normal form. 
The covariance relations (\ref{Eq:WickCovRel}) will give the symbols relative to 
\[
H'_{\rm cl} = H_{\rm cl}(\mathsf S_\mathbf{M} x) =  
%\tfrac{1}{2}\mathsf S_\mathbf{M} x \cdot{\bf M} \mathsf S_\mathbf{M} x =   
\tfrac{1}{2}x \cdot \Lambda_\mathbf{M} x =
\sum_{j=1}^{n} \frac{\mu_i}{2}(q_i^2 + p_i^2), 
\]
in terms of the product of 
$n$ symbols,  
one symbol in (\ref{ExE:WickSymbs}) for each frequency in $\Lambda_{\bf M}$. 
Similarly, the \hy{PF} will be product of 
$n$ \hy{PF}s in (\ref{ExE:PartFun}). 
This is nothing but unrevealing the normal-modes of a system of 
interacting oscillators, examples can be found in \cite{nicacioHC}.  

By another side, it is possible to show that the spectrum of a Hamiltonian
matrix $\mathsf J {\bf M}$ is equal to the one in (\ref{ExE:tw2}) 
if and only if ${\bf M} > 0$, see \cite{nicacioxx}.
Consequently, all thermal states generated by the \hy{QH} 
in (\ref{Eq:QQuadLinham}), or the symplectic matrix in (\ref{Eq:CompSymp}), 
are in category \hy{(E)} with ${\bf H} = {\bf M}$,
if and only if the Hessian of the \hy{QH} is positive-definite, 
${\bf M} >0$. 
In this case, using the covariance relation in (\ref{Eq:CovCompSymp}) 
with $\mathsf Q = \mathsf S_\mathbf{M}$, 
for $\mathsf S_\mathbf{M}$ in (\ref{ExE:tw1}), one finds
\[
\mathtt{S}'_{\beta} = 
{\rm e}^{-i\hbar \beta \mathsf J \Lambda_{\bf H}} = 
{\rm cosh}(\tfrac{1}{2} \hbar\beta \Lambda_{\bf H}) 
- i {\sf J} \,{\rm sinh}(\tfrac{1}{2} \hbar\beta \Lambda_{\bf H}), 
\]
where the last equality is obtained by a Taylor expansion of the exponential
and noting that $[\Lambda_{\bf H},{\sf J}] = 0$. From Eq.(\ref{Eq:Cayley}), 
the Cayley parametrization for $\mathtt{S}'_{\beta}$ reads
\[
\mathbf{C}_{{\mathtt S}'_{\beta}} = -i\,  
{\rm tgh}(\tfrac{1}{2} \hbar\beta \Lambda_{\bf H}).
\]
The covariance relation in (\ref{Eq:CayleyCov}) enables one to find the 
Cayley parametrization for the original Hamiltonian, {\it viz.}, 
\[
\mathbf{C}_{{\mathtt S}_{\beta}} = 
\mathsf S_\mathbf{M}^{-\top} 
\mathbf{C}_{{\mathtt S}'_{\beta}} \mathsf S_\mathbf{M}^{-1} = 
-i \, \mathsf S_\mathbf{M}^{-\top} 
{\rm tgh}(\tfrac{1}{2} \hbar\beta \Lambda_{\bf H})
\mathsf S_\mathbf{M}^{-1}.
\]
Since ${\rm tgh}(\tfrac{1}{2} \hbar\beta \Lambda_{\bf H}) >0$ for $\beta >0$, 
Eq.(\ref{Eq:SignCayleyCov}) shows that 
${\rm Im} \mathbf{C}_{{\mathtt S}_{\beta}} < 0$, 
and the symbols for the thermal operator in 
Eqs.(\ref{Eq:WeylSymThermal},\ref{Eq:WigSymThermal}) or the functions in (\ref{Eq:WigCharFunc}) are all Gaussians. 

Using the Williamson theorem for the matrix 
$\Lambda_{\bf H'} := {\rm tgh}(\tfrac{1}{2} \hbar\beta \Lambda_{\bf H})$, 
the matrix ${\rm Im} \mathbf{C}_{{\mathtt S}_{\beta}}$ 
[for above $\mathbf{C}_{{\mathtt S}_{\beta}}$] 
is able to produce any symmetric positive-definite matrix. 
In such a way, all Gaussian states can be reproduced 
by Eq.(\ref{Eq:WigCharFunc}) with above $\mathbf{C}_{{\mathtt S}_{\beta}}$. 
This also includes the pure states as the limit $\beta\to \infty$.

%%%%%%%%%%%%%%%%%%%%%%%%%%%%%%%%%%%%%%%%%%%%%%%%%%%%%%%%%%%%%%%%%%%%%%%%%%%%%%%%%%%%%%%%%
\subsection{Loxodromic Hamiltonian}                                %\label{Sec:sue} %%%%%
%%%%%%%%%%%%%%%%%%%%%%%%%%%%%%%%%%%%%%%%%%%%%%%%%%%%%%%%%%%%%%%%%%%%%%%%%%%%%%%%%%%%%%%%%
The loxodromic case can occur only in a phase space with at least four dimensions, 
what can be seen by the quartet structure of the eigenvalues in \hy{(L)}. 
Examples of systems where it appears are the ones in which a body subjected 
to an attractive potential also experiences torque forces, for instance, 
a spherical pendulum in which the symmetry axis itself is rotating with 
constant zenital angular velocity \cite{ozoriobook}.   
In these systems, centrifugal (centripetal) resultants  push (pull) 
the body away from (towards) an equilibrium configuration, 
which acting together with the attractive force, 
cause spiraling movements in phase space, see Eq.(\ref{ExL:SympEv}).

The Hamiltonian normal form of a system with two degrees of freedom 
in category \hy{(L)} is \cite{arnold}
\[%begin{equation}                                                                         \label{ExL:Ham}
H_{\rm cl} = \kappa (p_1 q_1 + p_2 q_2) + \omega \left(p_{2}q_{1} -  p_{1} q_{2}\right). 
\]%end{equation}
The Hessian and the corresponding Hamiltonian matrix are
\[
{\bf H} =       \begin{pmatrix}
                {\bf 0}_2 & {\bm \Omega} \\
                {\bm \Omega}^\top & {\bf 0}_2  
                \end{pmatrix},
\,\,                
\mathsf J {\bf H} = {\bm \Omega}^\top \oplus (-{\bm \Omega}),
\,\, 
{\bm \Omega} := \begin{pmatrix}
                 \kappa & -\omega \\
                 \omega & \kappa
                \end{pmatrix},  
\]
which generates the symplectic matrix
\begin{equation}                                                                         \label{ExL:SympEv}
\mathsf S_{t} = {\mathsf R_{t}} \, {\rm e}^{\kappa t} \oplus 
                {\mathsf R_{t}} \, {\rm e}^{-\kappa t} \in {\rm Sp}(4,\mathbb R),
\end{equation}
with 
\[
{\mathsf R_{t}} := \begin{pmatrix}
                   \cos(\omega t) & \sin(\omega t) \\
                   -\sin(\omega t) & \cos(\omega t)
                \end{pmatrix} \in {\rm Sp}(2,\mathbb R).
\]

The complex symplectic matrix in (\ref{Eq:CompSymp}) becomes
\[
\mathtt S_{\beta} = {\mathtt R_{\beta}} \, {\rm e}^{-i \hbar \kappa \beta t} \oplus 
                    {\mathtt R_{\beta}} \, {\rm e}^{i \hbar \kappa \beta t},
\]
with
\[
{\mathtt R_{\beta}} := \begin{pmatrix}
                       \cosh(\hbar\omega\beta  ) & -i\sinh(\hbar\omega\beta) \\
                      i\sinh(\hbar\omega\beta) & \cosh(\hbar\omega\beta)
                       \end{pmatrix}.
\]
The Cayley parametrization is 
\[
{\bf C}_{\mathtt S_{\beta}} = 
-i 
\begin{pmatrix}
{\bf 0}_4 & {\bf S} \\
{\bf S}^\top   & {\bf 0}_4  
\end{pmatrix}
\]
with 
\[
{\bf S} := \frac{
\begin{pmatrix}
 \sin (\kappa \beta  \hbar ) & -\sinh (\hbar \omega  \beta    ) \\
 \sinh (\hbar \omega  \beta    ) & \sin (\kappa \beta  \hbar ) \\
\end{pmatrix}}
{\cosh (\hbar \omega  \beta    )+\cos ( \hbar \kappa \beta )}
\]
and the four eigenvalues of ${\rm Im}{\bf C}_{\mathtt S_{\beta}}$ are  
\[
{\rm Spec}({\rm Im}{\bf C}_{\mathtt S_{\beta}}) =
\left\{
        \pm \frac{ \sqrt{ \cosh (2 \hbar \omega  \beta ) \pm 
                          \cos  (2  \hbar \kappa \beta )}            }
                 { \sqrt{2} (\cosh (\hbar \omega  \beta )+\cos ( \hbar \kappa \beta )) }
\right\}, 
\]
which are pairs of symmetric real numbers. Since 
\[
\det(\mathtt S_{\beta} \mp \mathsf I_{2n}) = 
4[\cos ( \hbar \kappa \beta ) \mp \cosh(\hbar \omega  \beta )]^2 ,   \\
\]
there is no Weyl representation only for $\beta = 0$, 
while the Wigner representation is well defined for any value $\beta \ge 0$. 
However, since the eigenvalues of ${\rm Im}{\bf C}_{\mathtt S_{\beta}}$ 
have different signs, these symbols are not related by the 
Fourier transformation (\ref{Eq:SympFour}). 
From (\ref{Eq:WCZlimit}), and since there are no divergences, 
$\nu^+_{\mathtt S_{\beta}} = 0, \, \forall \beta > 0$, 
and also $\nu^-_{\mathtt S_{\beta}} = 0$, 
due to the positivity of the \hy{PF} in (\ref{Eq:PartQuad}). 

%%%%%%%%%%%%%%%%%%%%%%%%%%%%%%%%%%%%%%%%%%%%%%%%%%%%%%%%%%%%%%%%%%%%%
%%%%%%%%%%%%%%%%%%%%%%%%%%%%%%%%%%%%%%%%%%%%%%%%%%%%%%%%%%%%%%%%%%%%%
%%%%%%%%%%%%%%%%%%%%%%%%%%%%%%%%%%%%%%%%%%%%%%%%%%%%%%%%%%%%%%%%%%%%%
\begin{figure}[b]
\includegraphics[width=7.5cm, trim=0 15 0 0]{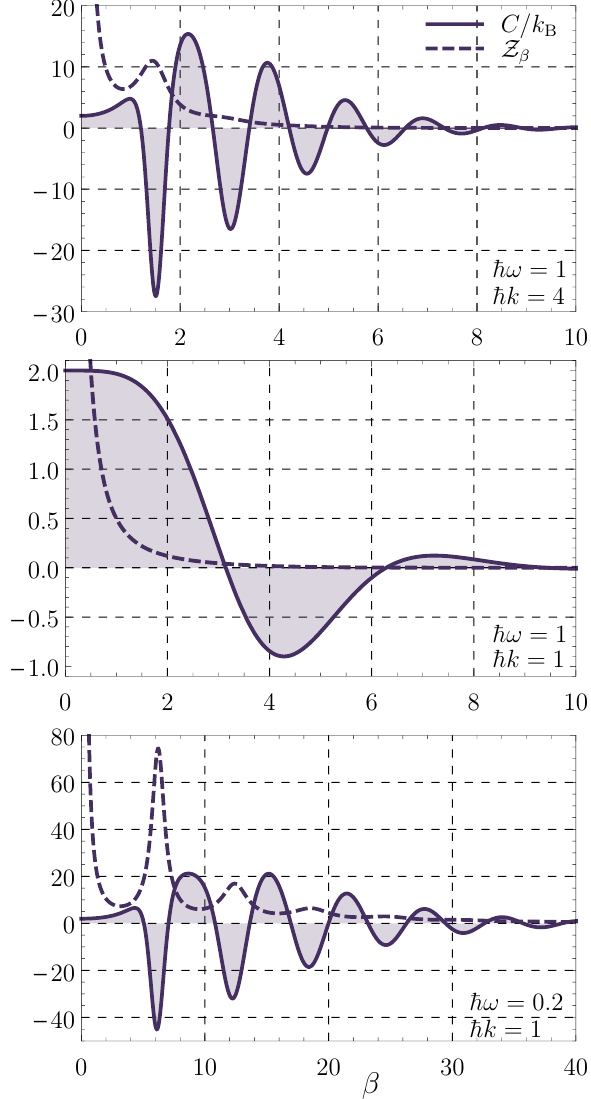}
\caption{ Partition function (dashed) and 
the Heat Capacity (continuous) 
of the Thermal State for the the two degrees of freedom 
Loxodromic Hamiltonian 
$H(x) = \kappa (p_1 q_1 + p_2 q_2) + 
\omega \left(p_{2}q_{1} -  p_{1} q_{2}\right)$ 
as a function of the inverse temperature $\beta$. 
Top: $\hbar \omega = 1$ and $\hbar \kappa = 4$.
Middle: $\hbar \omega = 1$ and $\hbar \kappa = 1$. 
Bottom: $\hbar \omega = 0.2$ and $\hbar \kappa = 1$.}                                        \label{Fig:L}
\end{figure}
%%%%%%%%%%%%%%%%%%%%%%%%%%%%%%%%%%%%%%%%%%%%%%%%%%%%%%%%%%%%%%%%%%%%%
%%%%%%%%%%%%%%%%%%%%%%%%%%%%%%%%%%%%%%%%%%%%%%%%%%%%%%%%%%%%%%%%%%%%%
%%%%%%%%%%%%%%%%%%%%%%%%%%%%%%%%%%%%%%%%%%%%%%%%%%%%%%%%%%%%%%%%%%%%%

The \hy{PF} in (\ref{Eq:PartQuad}) becomes 
\[
\mathcal Z_\beta = 
\frac{1}{2} [ \cosh(\hbar \omega  \beta ) - \cos ( \hbar \kappa \beta )]^{-1},  
\]
and the Heat capacity in (\ref{Eq:HeatCap}), 
\[
\begin{aligned}
C &= k_{\rm B}\hbar^2 (\omega^2 - \kappa^2) \beta^2 
      \frac{  
              \cosh (\beta  \omega  \hbar ) \cos (\beta  \kappa \hbar ) - 1 }
           { [\cos (\beta  \kappa \hbar )-\cosh (\beta  \omega  \hbar )]^2    } \\
    &+ 2 \kappa_{\rm B} \hbar ^2  \kappa \omega \beta^2  
    \frac{  \sinh (\beta  \omega  \hbar ) \sin (\beta  \kappa \hbar )}
           { [\cos (\beta  \kappa \hbar )-\cosh (\beta  \omega  \hbar )]^2}.
\end{aligned}
\]
Both are potted in Fig.\ref{Fig:L} and from there it is possible to observe that 
the parameter $\kappa$ controls the oscillatory behavior, 
while $\omega$ controls the amplitude. In agreement with (\ref{Eq:EquipTeo}),  
$\lim_{\beta \to 0} C = 2 k_{\rm B}$.
The thermodynamical instability of this kind of systems, 
associated to negative values of the heat capacity \cite{statphys}, 
is a property of the whole class \hy{(L)}, 
since the heat capacity is related to the concavity of the \hy{PF}, 
see Eq.(\ref{Eq:HeatCap}), and the \hy{PF} only depends on the eigenvalues 
of the Hamiltonian matrix $\mathsf J \bf H$.   

%%%%%%%%%%%%%%%%%%%%%%%%%%%%%%%%%%%%%%%%%%%%%%%%%%%%%%%%%%%%%%%%%%%%%%%%%%%%%%%%%%%%%%%%%
%%%%%%%%%%%%%%%%%%%%%%%%%%%%%%%%%%%%%%%%%%%%%%%%%%%%%%%%%%%%%%%%%%%%%%%%%%%%%%%%%%%%%%%%%
\section{Conclusions}                                                    \label{Sec:conc}
%%%%%%%%%%%%%%%%%%%%%%%%%%%%%%%%%%%%%%%%%%%%%%%%%%%%%%%%%%%%%%%%%%%%%%%%%%%%%%%%%%%%%%%%%
%%%%%%%%%%%%%%%%%%%%%%%%%%%%%%%%%%%%%%%%%%%%%%%%%%%%%%%%%%%%%%%%%%%%%%%%%%%%%%%%%%%%%%%%%
% The main question addressed in this work was the determination of Wigner-Wieyl symbols 
% for canonical equilibrium states of \hy{QH}s, 
% which was achieved by an application of a Wick rotation. 
% The result were supported on the well developed theory 
% for the unitary operator generated by \hy{QH}s. 
%

The class of Hamiltonian systems is extensively broader than 
the quadratic case, notwithstanding, as it is learnt from dynamical systems, 
much of the system evolution is grasped on its fixed points 
whose nature determines locally the system behavior through a linearization, 
rising a \hy{QH}. 
Under this perspective, 
it is opportune to conclude with some remarks on missing points 
and possible extensions.

The prototype of a thermodynamical system is the ideal gas \cite{statphys}, 
which is a system described by a \hy{QH} in class \hy{(P)} 
and with Hessian as that one in (\ref{ExP:Hess}). 
As a mechanical system, a canonical transformation 
of phase space coordinates would not change its physical properties.
Nevertheless, its \hy{PF} (\ref{ExP:partfunc}) and all of 
its thermodynamical properties definitively are not subjected to the 
covariance rules in Sec.\ref{Sec:cos}. 
Technically speaking, 
this is due to the performed truncation of an integral in (\ref{ExP:partfunc}).
Physically, the reason relies on the imposition of a container with finite volume, 
which appears as a ``contour-condition", or a non-holonomic constraint, 
whence not canonically covariant.

The divergences of the \hy{PF} for the Hamiltonian systems 
in category \hy{(H)} are not tamed by such kind of constrain, 
see Eq.(\ref{ExH:parttrunc}), 
showing that these are related to distinct properties of a thermodynamical system. 
At this point, the advantages of the Wigner-Weyl representations become clear: 
the analytical expression of the \hy{PF} is obtained 
through the Weyl representation,  
despite the non-convergence of the integral of the Wigner symbol (\ref{ExH:WigWickSymb}).   
Such kind of divergence, a consequence of the sum in the \hy{PF} 
for a continuous and unbound energy levels of the hyperbolic Hamiltonian \cite{bollini},
is what happens in scattering problems, 
which is exemplified at the end of Sec.\ref{Sec:hh}. 

A diverging (or not converging) \hy{PF} in principle is a pathology for the 
statistical treatment of physical systems. 
Quoting Gibbs \cite{Gibbs}, 
``we shall always suppose the (...) [partition function] to have a finite value, 
as otherwise the coefficient of probability vanishes, and the law of distribution 
becomes illusory. 
This will exclude certain cases (...) for instance, cases in which the system 
or parts of it can be distributed in unlimited space (...)". 
If, by one side, 
the non-convergence for the category \hy{(P)} is amended by the constrain imposition, 
category \hy{(H)} [and also \hy{(L)}], as it is, 
can play a privileged role in the development 
of a full dynamical background to statistical physics, since these Hamiltonians are 
the ones which possesses positive Lyapunov exponents, which are responsible for the 
phenomena of mixing in classical chaotic Hamiltonians \cite{ozorio1998,gutzwiller}. 
The relation of mixing and relaxation to
equilibrium is a question posed in \cite{casati} 
and analyzed for chaotic billiards. 
It remains an open question for generic non-linear Hamiltonians.  

The other prescription of exclusion in \cite{Gibbs} is when 
``the energy can decrease without limit, 
as when the system contains material points which attract 
one another inversely as the squares of their distances". 
While classically, it is impossible to write a \hy{PF} for 
gravitational or Coulomb interacting particles due to the boundlessness of system energy, 
the quantum \hy{PF} of such systems also diverges, 
but due to the energy level spacing structure \cite{strickler}. 
Despite such Hamiltonians does not have any fixed point, 
since they are such that $\partial H / \partial x \ne 0$,    
the dynamics of the radial coordinate \cite{arnold} lives in a bifurcative scenario 
between \hy{(E)} and \hy{(H)}: negative energies originate ellipses, 
zero energy trajectories are parabolas, and positive energy are hyperbolas 
in configuration space. 
This bifurcative behavior and its relation with the divergence of partition
function, both classical and quantum, should be clarified in a future work.

From the case of \hy{(H)}, the thermal capacity diverges when the temperature 
approaches zero, which is also a consequence of the boundlessness of the system, 
since at this temperature limit it does not attain a ground state, a prerequisite of the 
Nernst principle \cite{statphys}. 
More interesting is the category \hy{(L)}, where the system has negative heat capacity
for certain ranges of temperatures, see Fig.\ref{Fig:L}.  
However, be in this category is only a sufficient condition for a system to present
this behavior. 
The system presented in \cite{staniscia} is a collection of two dimensional rotors 
(four dimensional phase-space).  
The Hamiltonian of two interacting rotors has a fixed point associated simultaneously 
with a pair of eigenvalues in \hy{(P)} and another pair in \hy{(E)}. 
This seems to show that a combination of different categories can generate the negative
specific heat, which is out of the scope of this work, however should be investigated.
Another example where there is a combination of categories for the same fixed point is 
\cite{posch} and again the system has a negative heat capacity. %
Both systems in \cite{staniscia,posch} 
have long-range interactions and one can not expect that the local approximation 
developed in Sec.\ref{Sec:gh} to work. 
A better method of approximation which takes into account the influence 
of multiple or hybrid fixed points spread in phase space should be developed.  

Few months after the submission of this work, 
the reference \cite{ozorio2} dealing with similar questions  
have appeared in arXiv.  

%%%%%%%%%%%%%%%%%%%%%%%%%%%%%%%%%%%%%%%%%%%%%%%%%%%%%%%%%%%%%%%%%%%%%%%%%%%%%%%%%%%%%%%%%
%%%%%%%%%%%%%%%%%%%%%%%%%%%%%%%%%%%%%%%%%%%%%%%%%%%%%%%%%%%%%%%%%%%%%%%%%%%%%%%%%%%%%%%%%
\acknowledgments         %%%%%%%%%%%%%%%%%%%%%%%%%%%%%%%%%%%%%%%%%%%%%%%%%%%%%%%%%%%%%%%%
%%%%%%%%%%%%%%%%%%%%%%%%%%%%%%%%%%%%%%%%%%%%%%%%%%%%%%%%%%%%%%%%%%%%%%%%%%%%%%%%%%%%%%%%%
%%%%%%%%%%%%%%%%%%%%%%%%%%%%%%%%%%%%%%%%%%%%%%%%%%%%%%%%%%%%%%%%%%%%%%%%%%%%%%%%%%%%%%%%%
%
\noindent The author acknowledges the warm hospitality of 
Profs. H.G. Feichtinger and M. de Gosson from NuHAG -- 
Universität Wien. 
The author is a member of the Brazilian National Institute of Science 
and Technology for Quantum Information [CNPq INCT-IQ (465469/2014-0)] 
and also acknowledges CAPES [PrInt2019 (88887.468382/2019-00)].

%%%%%%%%%%%%%%%%%%%%%%%%%%%%%%%%%%%%%%%%%%%%%%%%%%%%%%%%%%%%%%%%%%%%%%%%%%%%%%%%%%%%%%%%%

\end{document}